\documentclass{article}

\usepackage{microtype}
\usepackage{graphicx}
\usepackage{subfigure}
\usepackage{booktabs} 

\usepackage{hyperref}

\usepackage[accepted]{icml2025}

\usepackage{amsmath}
\usepackage{amssymb}
\usepackage{mathtools}
\usepackage{amsthm}

\usepackage[capitalize,noabbrev]{cleveref}
\usepackage[T1]{fontenc}

\theoremstyle{plain}
\newtheorem{theorem}{Theorem}[section]

\theoremstyle{definition}
\newtheorem{definition}[theorem]{Definition}

\theoremstyle{remark}

\usepackage[textsize=tiny]{todonotes}
\usepackage[absolute]{textpos}

\icmltitlerunning{
In-Context Adaptation to Concept Drift for Learned Database Operations}

\usepackage{mathtools}      

\usepackage{xspace}
\newcommand{\name}{FLAIR\xspace}

\usepackage{xcolor} 
\usepackage{colortbl}
\usepackage{amsopn}

\usepackage{makecell}
\usepackage{color}
\newcommand{\highlight}[1]{\textbf{#1}}
\newcommand{\blue}[1]{{\color{black}#1}}

\newcommand{\update}[1]{{\color{black}#1}}
\newcommand{\term}[1]{{\color{black}#1}}
\newcommand{\green}[1]{{\color{black}#1}}

\newcommand{\icmlblue}[1]{{\color{black}#1}}
\newcommand{\ignore}[1]{}

\usepackage{enumitem}
\newcommand{\squishlist}
{
	\begin{list}{$\bullet$}
		{
			\setlength{\itemsep}{0pt}
			\setlength{\parsep}{3pt}
			\setlength{\topsep}{3pt}
			\setlength{\partopsep}{0pt}
			\setlength{\leftmargin}{1.5em}
			\setlength{\labelwidth}{1em}
			\setlength{\labelsep}{0.5em}
		}
	}
	
	\newcommand{\squishend}
	{
	\end{list}
}

\usepackage{multirow}
\usepackage{arydshln}
\usepackage{caption}
\usepackage{subcaption}
\usepackage{makecell}           
\usepackage{booktabs}  
\newsavebox{\measurebox}

\usepackage{multicol}

\usepackage{url}

\usepackage{dsfont}

\usepackage{listings}

\usepackage{colortbl}
\definecolor{basegray}{gray}{.9}
\definecolor{myblue}{cmyk}{.3,0,0,0}
\definecolor{baseblue}{cmyk}{.1,0,0,0}

\definecolor{baselgray}{cmyk}{.02,0.01,0.01,0.01}
\definecolor{mylblue}{cmyk}{.02,0.01,0.01,0.01}

\usepackage{tikz}
\usetikzlibrary{tikzmark}

\usepackage{amsmath}

\usepackage{enumitem}
\usepackage{mathabx}
\usepackage{graphicx}
\usepackage{multicol}

\begin{document}

\twocolumn[
\icmltitle{
In-Context Adaptation to Concept Drift for Learned Database Operations}



\icmlsetsymbol{equal}{*}

\begin{icmlauthorlist}
\icmlauthor{Jiaqi Zhu}{bit}
\icmlauthor{Shaofeng Cai}{equal,nus}
\icmlauthor{Yanyan Shen}{sj}
\icmlauthor{Gang Chen}{zju}
\icmlauthor{Fang Deng}{bit}
\icmlauthor{Beng Chin Ooi}{nus,zju}
\end{icmlauthorlist}

\begin{center}
$^1$ Beijing Institute of Technology,
  $^2$ National University of Singapore,\\
  $^3$ Shanghai Jiao Tong University,
  $^4$ Zhejiang University\\
  \{jiaqi\_zhu, dengfang\}@bit.edu.cn, shaofeng@comp.nus.edu.sg, shenyy@sjtu.edu.cn, \{cg, ooibc\}@zju.edu.cn
\end{center}




\icmlkeywords{Concept Drift, Learned Database Operation, Online Adaptation}
\vskip 0.3in
]




\ifnum\value{page}=1
\begin{textblock*}{\columnwidth}(20mm,256mm)
\noindent\small\textit{*Corresponding author\\
Published as a conference paper at ICML 2025}
\end{textblock*}
\fi

\begin{abstract}

\ignore{
\icmlblue{
With the growing support of machine learning for in-database analytics,
significant performance benefits have been realized in enhancing database operations.
However, the continual evolution of database content leads to \textit{concept drift}, posing persistent challenges for learned database operations in maintaining accuracy. 
Although methods like transfer learning and active learning are employed to alleviate such a problem, they are resource-intensive due to the necessity for frequent data recollection and model retraining.}
}

Machine learning has demonstrated transformative potential for database operations, such as query optimization and in-database data analytics.
However, dynamic database environments, characterized by frequent updates and evolving data distributions, introduce \textit{concept drift}, which leads to performance degradation for learned models and limits their practical applicability.
Addressing this challenge requires efficient frameworks capable of adapting to shifting concepts while minimizing the overhead of retraining or fine-tuning.

In this paper, we propose \name, an online adaptation framework that introduces a new paradigm called \textit{in-context adaptation} for learned database operations.
\name leverages the inherent property of data systems, i.e., immediate availability of execution results for predictions, to enable dynamic context construction.
By formalizing adaptation as $f:(\mathbf{x} \,| \,\mathcal{C}_t) \to \mathbf{y}$, with $\mathcal{C}_t$ representing a \textit{dynamic context memory}, \name delivers predictions aligned with the current concept, eliminating the need for runtime parameter optimization.
To achieve this, \name integrates two key modules: a Task Featurization Module for encoding task-specific features into standardized representations, and a Dynamic Decision Engine, pre-trained via Bayesian meta-training, to adapt seamlessly using contextual information at runtime.
Extensive experiments across key database tasks demonstrate that \name outperforms state-of-the-art baselines, achieving up to $5.2\times$ faster adaptation and reducing error by 22.5\% for cardinality estimation.

\end{abstract}
\vspace{-5mm}
\section{Introduction}\label{sec:introduction}

Data systems are increasingly integrating machine learning functionalities to enhance performance and usability, marking a paradigm shift in how data is managed and processed in databases~\cite{ooi2024neurdb,
mcgregor2021preventing,li2021ai}.
The integration has transformed key database operations such as query optimization, indexing, and workload forecasting into more precise, efficient, and adaptive processes~\cite{zhang2024making,kurmanji2023detect,anneser2023autosteer,ferragina2020learned}.

Despite these advancements, learned database operations face a persistent challenge: \textit{concept drift}.
Databases are inherently dynamic, undergoing frequent insert, delete, and update operations that result in shifts in data distributions and evolving input-output relationships over time~\cite{zeighami2024theoretical}.
These drifts, often subtle but cumulative, can alter the patterns and mappings that traditional machine learning models rely upon, rendering their assumptions of static distributions invalid.
This phenomenon requires adaptive methods for maintaining predictive accuracy in dynamic database environments.

\begin{figure*}[t]
    \centering 
\includegraphics[width=1\linewidth]{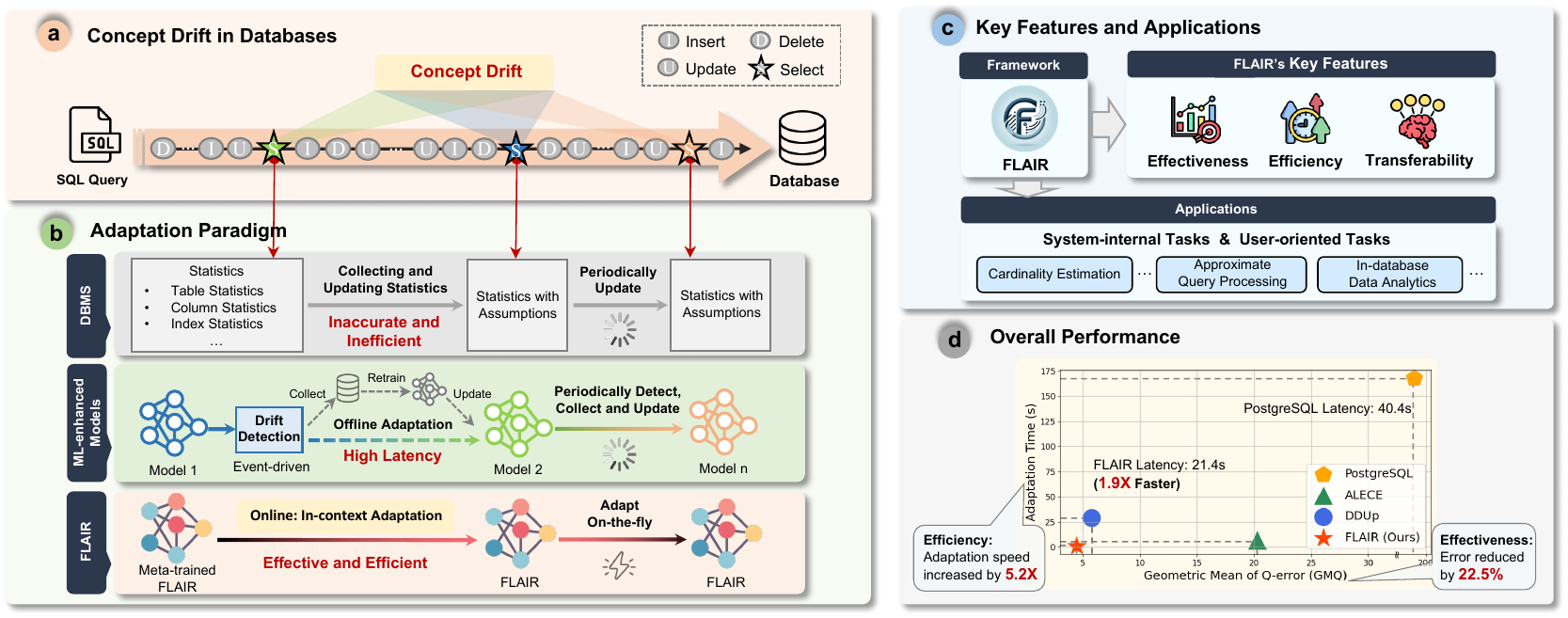} \vspace{-6mm}
    \caption{\name in a nutshell. 
    (a) An example of concept drifts in a dynamic database setting.
    (b) 
    Adaptation paradigm for handling concept drifts in \name and two conventional approaches.
    (c)  
    Key features and applications of \name.
    (d) A preview comparison of \name 
    against
    PostgreSQL 
    and state-of-the-art approaches for handling dynamic databases.
    }
    \label{fig:intro}
    \vspace{-5mm}
\end{figure*}

Traditional \textit{reactive training-based adaptation} approaches to handling concept drift, such as
transfer learning~\cite{jain2023data,kurmanji2023detect, kurmanji2024machine}, active learning~\cite{ma2020active,li2022warper}, and multi-task learning~\cite{kollias2024distribution,wu2021unified},
come with significant drawbacks in learned database operations.
As illustrated in Figure~\ref{fig:intro}, delays and costs in post-deployment data collection and model updates, and reliance on static mappings limit their practicality in dynamic database environments~\cite{kurmanji2024machine,li2022warper}.
In addition, they process each input independently.
The negligence of inter-query dependencies and shared contextual information in databases results in poor modeling of database operations.
Addressing these limitations raises two critical challenges:
(1) \textit{How can we support one-the-fly adaptation to constantly evolving data without incurring the overhead of frequent retraining or fine-tuning in databases?}
(2) \textit{How can we dynamically inject contextual information into the modeling process to achieve context-aware prediction for learned database operations?}

\blue{
}

To address these challenges, we introduce \name, an e\underline{F}ficient and effective on\underline{L}ine \underline{A}daptat\underline{I}on f\underline{R}amework that establishes a new adaptation paradigm for learned database operations.
\name is built on a unique property of database operations: the immediate availability of execution results for predictions in the database.
These results, serving as ground-truth labels, provide real-time feedback that enables seamless adaptation.
\name leverages this property to dynamically adapt to evolving concepts using such \textit{contextual cues} from databases.
Formally, \name models the mapping as $f:(\mathbf{x} \,| \,\mathcal{C}_t) \to \mathbf{y}$, where $\mathbf{x}$ denotes the input query, $\mathcal{C}_t$ is the current context consisting of recent pairs of queries and their execution results, and $\mathbf{y}$ is the predicted output.

To achieve \textit{in-context adaptation} for learned database operations, \name introduces two cascaded modules: the \textit{task featurization module} (TFM) and the \textit{dynamic decision engine} (DDE).
The TFM encodes database operations into standardized task representations, extracting informative features and producing a unified, structured input format.
This ensures consistency and efficiency across diverse tasks within databases.
The \textit{dynamic decision engine} functions as the core of \name, delivering predictions that can adapt to evolving concepts.
To this end, we introduce a Bayesian meta-training mechanism that utilizes synthetic prior distributions to pretrain \name with a comprehensive knowledge base, pre-adapting it to handle diverse and dynamic scenarios.
Unlike traditional reactive approaches, \name eliminates the need for compute-intensive parameter optimization after deployment.
To the best of our knowledge, \name is the first framework to enable on-the-fly and context-aware adaptation in dynamic data systems.

We summarize our main contributions as follows:
\begin{itemize}[itemsep=0mm,leftmargin=4mm]
\vspace{-4mm}
    \item
    \update{We propose a novel in-context adaptation framework \name, 
    designed to address the persistent challenge of concept drift in dynamic data systems with high efficiency and effectiveness.
    }
    
    \item
    \ignore{\update{We establish a new adaptation paradigm} by equipping \name with Bayesian meta-training to master learning from dynamic distributions.  
    By querying the meta-trained model with contextual information, \name provides accurate predictions guided by the context.}
    
    \name introduces Bayesian meta-training that enables robust and transferable learning from dynamic distributions, thus eliminating the need for costly parameter retraining or fine-tuning after deployment.

    
    \item
    \name is designed as a task-agnostic framework that enhances a wide range of learned database operations.
    These include system-internal tasks such as cardinality estimation, and user-oriented applications like approximate query processing and in-database data analytics.
    
    
    \item 
    Extensive experiments show \name's superior performance
    in dynamic databases, achieving a $5.2\times$ speedup in adaptation and a 22.5\% reduction in GMQ error for cardinality estimation.
    Furthermore, by integrating \name with PostgreSQL, we achieve up to a $1.9\times$ improvement in query execution efficiency.
    
    
\end{itemize}


\section{Preliminaries}
\label{sec:preliminary}

\noindent
\highlight{Problem Formulation.}
Consider a database $\mathrm{D}$ consisting of a set of relations (tables) $\{\mathbf{R_1},...,\mathbf{R_N}\}$. 
Each relation $\mathbf{R_i}$ has $n_i$ attribute fields (columns), $\mathbf{R_i}=(\mathbf{a_{1}^{i}},...,\mathbf{a_{n_i}^{i}})$, where the attributes correspond to either categorical or numerical features in prediction.
\update{In this paper, we focus on select-project-join (SPJ) queries 
executed alongside a mix of insert, delete, and update operations.
The challenge addressed is \textit{concept drift}, an intrinsic property of databases, described as a shift in the relationship between queries and their corresponding predictive outputs over time.
}
\begin{definition}[Concept Drift in Databases] \vspace{-1mm}
\label{def:concep_drift_db} 
\blue{Let $\mathrm{Q}=\{\mathbf{x_1},\mathbf{x_2},\cdots\}$ represents a sequence of \term{input queries} and $\mathrm{Y}=\{\mathbf{y_1},\mathbf{y_2},\cdots\}$ denote the corresponding \term{output predictions}, \update{e.g., estimated row counts in cardinality estimation}.
Concept drift occurs at time $t$ if the joint probability distribution 
$P_t\left(\mathbf{x},\mathbf{y} \right)$
changes from $P_t\left(\mathbf{x},\mathbf{y} \right)$ to $P_{t+1}\left(\mathbf{x},\mathbf{y} \right)$, such that
$P_t\left(\mathbf{x},\mathbf{y} \right) \ne P_{t+1}\left( \mathbf{x}, \mathbf{y} \right)$.}
\end{definition}\vspace{-2mm}

In concept drift, the change in the joint probability distribution $P\left( \mathbf{x}, \mathbf{y} \right)=P(\mathbf{x})P(\mathbf{y} |\mathbf{x})$ may come from shifts in $P(\mathbf{x})$ (\textit{covariate shift}) or 
$P(\mathbf{y} |\mathbf{x})$ (\textit{real shift}).
Database updates, especially frequent insert, delete and update operations, typically induce shifts in $P(\mathbf{y} |\mathbf{x})$, showing the dynamic nature of the data systems.
\update{
While individual updates might only marginally affect the underlying distribution, cumulative changes can significantly alter \term{query-prediction relationships}.
For example, in an e-commerce database, incremental updates, such as new product additions, customer preference changes, or promotional campaigns, can lead to significant concept drift in product recommendation.

}

\icmlblue{
\noindent
\highlight{Learned Database Operations.}}
Learned database operations employ machine learning models to enhance specific tasks in databases, such as cardinality estimation and approximate query processing.
Let $\mathcal{M}_{D}(\cdot; \Theta)$ denote a prediction model parameterized by $\Theta$ in a database $\mathrm{D}$.
$\mathcal{M}_{D}(\mathbf{x}; \Theta)$ takes a query $\mathbf{x}$ as input and makes a prediction, e.g., the number of rows matching $\mathbf{x}$ for cardinality estimation.

However, a model becomes stale when concept drift occurs.
Formally, the model $\mathcal{M}_{D_t}( \mathbf{x}; \Theta_t) $ trained on data $D_t$ becomes ineffective at time $t+\Delta t$, if 
$P_t \left(\mathbf{x},\mathbf{y} \right) \ne P_{t+\Delta t}\left( \mathbf{x}, \mathbf{y} \right)$.
Traditional approaches require periodic data recollection and model retraining to maintain accuracy.
This incurs high costs.
Our objective is to ensure that the model $\mathcal{M}_{D_t}( \mathbf{x}; \Theta_t)$ can be efficiently and effectively adapted to evolving data distributions with these resource-intensive processes in database environments.

\begin{figure}[t]
    \centering
\includegraphics[width=1\linewidth]{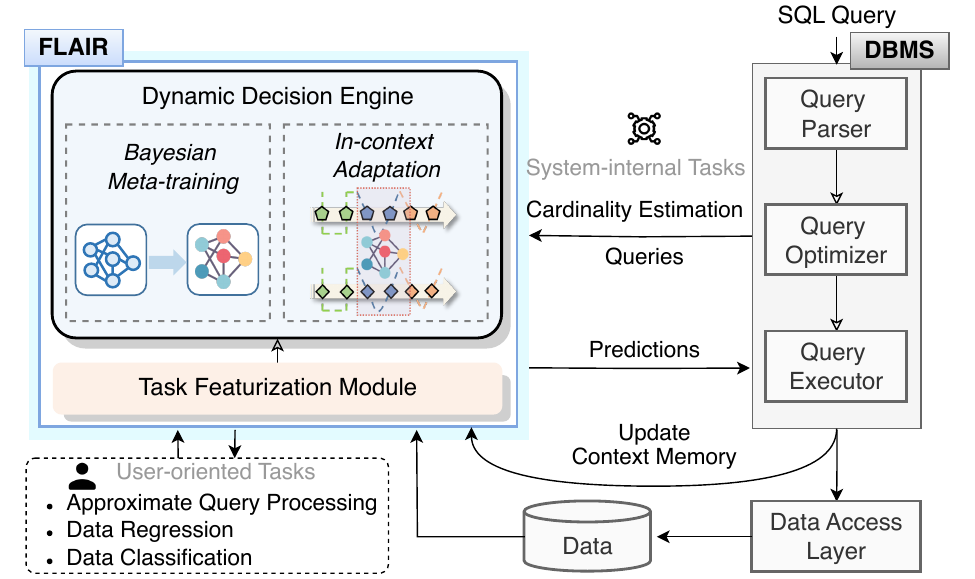}  \vspace{-3mm}
    \caption{\name for dynamic data systems. }
    \label{fig:framework}
    \vspace{-4mm}
\end{figure}

\icmlblue{
\noindent
\highlight{In-context Learning with Foundation Models.}}
Foundation models have seen rapid advancements in capability and scope~\cite{radford2019language,raffel2020exploring,brown2020language,achiam2023gpt}, which give rise to a transformative paradigm called \textit{in-context learning} (ICL).
ICL embeds context into the model input, and leverages foundation models' broad learned representations to make predictions based on limited contextual examples, thus bypassing the need for parameter updates after deployment.
This paradigm drastically cuts compute demands and facilitates various applications ~\cite{sun2022black,dong2022survey}.
A notable application for tabular data is \textit{Prior-data Fitted Networks} (PFNs)~\cite{muller2021transformers,hollmann2022tabpfn,helli2024drift}, which are pre-trained on synthetic datasets sampled from pre-defined priors.
This enables PFNs to pre-adapt to dynamic environments by effectively modeling uncertainties and various distributions, making PFNs suitable for scenarios with frequent updates and concept drift.
In this paper, we aim to utilize real-time feedback from database environments and explore how to support in-context adaptation for learned database operations.

\icmlblue{\section{\name for Dynamic Data Systems}}
\label{sec:methodology}
As illustrated in Figure~\ref{fig:framework}, \name introduces a dual-module architecture that addresses concept drift in dynamic databases.
First, to provide a unified interface across different tasks, the \term{Task Featurization Module (TFM)} extracts task-specific features from database operation for the subsequent modeling.
Second, the \term{Dynamic Decision Engine (DDE)} is pre-trained via Bayesian meta-training on dynamic distributions of tasks, pre-adapting it to diverse tasks encountered during inference.
After meta-training, DDE utilizes the real-time
feedback from databases as the latest \term{contextual information} to dynamically adapt to the current task.
The workflow of \name $\mathcal{M}_{F}$ is outlined as:
\begin{align}
\textstyle
\label{eq:model_workflow}
\mathcal{M}_{F}(\mathbf{x}; \Theta_\mathcal{T},\Theta_\mathcal{D})=\mathcal{M}_{DDE}( \mathcal{M}_{TFM}(\mathbf{x}; \Theta_\mathcal{T}); \Theta_\mathcal{D}),
\end{align}
which comprises two cascading modules, the TFM $\mathcal{M}_{TFM}$ and the DDE $\mathcal{M}_{DDE}$ parameterized by $\Theta_\mathcal{D}$ and $\Theta_\mathcal{T}$, respectively.
We introduce the technical details below.


\subsection{Task Featurization Module}\label{sec:attention-based_modeling}
The TFM is designed to standardize database operations into structured inputs for downstream modeling.
It first encodes data and queries of database operations into \textit{\term{data vectors}} and a \textit{\term{query vector}} respectively, and then extracts a \textit{\term{task vector}} via cross-attention that integrates their interactions.
\subsubsection{Data and Query Encoding}

\noindent \textbf{Data Encoding.}
Each attribute (i.e., column) in the database is represented as a histogram, which captures its distribution.
Formally, for an attribute $\mathbf{a_{n}^{i}}$ in relation $\mathbf{R_i}$, the histogram $\mathbf{x}_{n}^{i} = [x_1, \cdots, x_\delta]$ uses $\delta$ bins to discretize the range of the attribute.
After scaling to $[0,1]$, these histograms are aggregated to form comprehensive \term{data vectors $\mathrm{X_D}$} of dimension $\delta \times \sum_{i=1}^{N}{n_i}$, where $N$ is the total number of relations, and $n_i$ is the number of attributes in relation $\mathbf{R_i}$.

\noindent \textbf{Query Encoding.}
Queries are represented as vectors capturing structural and conditional information.
Join predicates, e.g., $\mathbf{R_i}\mathbf{a_{n_i}^{i}} = \mathbf{R_j}\mathbf{a_{n_j}^{j}}$, are encoded into binary vectors $\mathbf{q}_J$ via one-hot encoding, while filter predicates, e.g.,  $\mathbf{R_i}\mathbf{a_{n_i}^{i}} \text{ op } \Omega$ with $\text{op} \in \{<, \leq, \geq, >, =\}$ being the comparison operators and $\mho$ the condition value, are encoded into boundary vectors $\mathbf{q}_F$.
The final \textit{\term{query vector}} $\mathbf{q}_\mathcal{Q}=<\mathbf{q}_J,\mathbf{q}_F>$ concatenates these encodings.

\subsubsection{Task Featurization}
To derive the task vector, we adopt a lightweight transformer~\cite{vaswani2017attention} architecture following~\cite{li2023alece}, which employs hybrid attention mechanisms to extract deep latent features.
The task featurization process starts with a \textit{data modeling phase,}
where data vectors $\mathrm{X_D}$ are processed through a series of Multi-head Self-attention (MHSA) layers, interleaved with Feed-forward Network (FFN), Layer Normalization (LN), and residual connections.
This is to capture implicit joint distributions and complex dependencies among attributes within $\mathrm{X_D}$:
\begin{align}
\label{eq:selfatt}
\mathbf{\hat{Z}}^l &= {\rm MHSA}({\rm LN}(\mathbf{Z}^{l-1})) + \mathbf{Z}^{l-1} \\
\mathbf{Z}^l &= {\rm FFN}({\rm LN}(\mathbf{\hat{Z}}^l))+\mathbf{\hat{Z}}^l
\end{align}
\noindent where MHSA operations are formulated as: 
\begin{small}
\begin{flalign}
\raisetag{16pt}
&\scalebox{0.97}{$\mathbf{Q}^{l,m}=\mathbf{Z}^{l-1}\mathbf{W}_q^{l,m},\mathbf{K}^{l,m}=\mathbf{Z}^{l-1}\mathbf{W}_k^{l,m},\mathbf{V}^{l,m}=\mathbf{Z}^{l-1}\mathbf{W}_v^{l,m}$}&&\\ 
&\mathbf{Z}^{l,m} = {\rm softmax}(\frac{\mathbf{Q}^{l,m}(\mathbf{K}^{l,m})^T}{\sqrt{d_k}})\mathbf{V}^{l, m}, \,\, m=1,\cdots,M&& \\
&\mathbf{Z}^{l} = {\rm concat}(\mathbf{Z}^{l,1},\cdots,\mathbf{Z}^{l,M})\mathbf{W}_o^l&&
\label{eq:att_value}
\end{flalign}
\end{small}

\vspace{-2mm}
\noindent
where $\mathbf{Z}^{0}$ is composed of data vectors from $\mathrm{X_D}$,
and $M$ is the number of attention heads.
$\mathbf{Q}^{l,m}$, $\mathbf{K}^{l,m}$, and $\mathbf{V}^{l,m}$ denote the \textit{query}, \textit{key}, and \textit{value} of the $m$-th head in the $l$-th layer, obtained via transformation matrices $\mathbf{W}_q^{l,m}$, $\mathbf{W}_k^{l,m}$, and $\mathbf{W}_v^{l,m}$, respectively.
$\mathbf{Z}^{l}$ is the output of the $l$-th layer, and  $\mathbf{W}_o^l$ is the output transformation matrix.

In the subsequent \textit{interaction modeling phase}, the output of the data modeling phase $\mathbf{Z}_\mathcal{O}$ is further refined via the Multi-head Cross-attention (MHCA) mechanism.
Unlike MHSA, $\mathbf{Z}_\mathcal{O}$ serves dual roles as both the \textit{keys} and \textit{values}, while the query vector $\mathbf{q}_\mathcal{Q}$ acts as the \textit{query} in MHCA.
The query vector $\mathbf{q}_\mathcal{Q}$ interacts with every vector in $\mathbf{Z}_\mathcal{O}$ through key and value transformations, allowing TFM to dynamically focus on the features in $\mathbf{Z}_\mathcal{O}$ pertinent to the query.
For each attention head $m$ in MHCA, we have:
\begin{small}
\begin{align}
&\mathbf{z}^m = {\rm softmax}(\frac{\mathbf{q}_\mathcal{Q}(\mathbf{Z}_\mathcal{O}\mathbf{W}_k^m)^T}{\sqrt{d_k}})(\mathbf{Z}_\mathcal{O}\mathbf{W}_v^m).
\label{eq:crossatt}
\end{align}
\end{small}

\vspace{-2mm}
\noindent
The final task vector $\mathbf{z}_\mathcal{T}$ is obtained by further processing the MHCA output through an FFN layer followed by LN with residual connections.
In this way, 
the task vector $\mathbf{z}_\mathcal{T}$ contains task-specific information of both data attribute relations and query conditions, providing comprehensive task representations for the subsequent modeling in the DDE.

\begin{figure}[t]
    \centering
    \vspace{-1mm}
\includegraphics[width=0.95\linewidth]{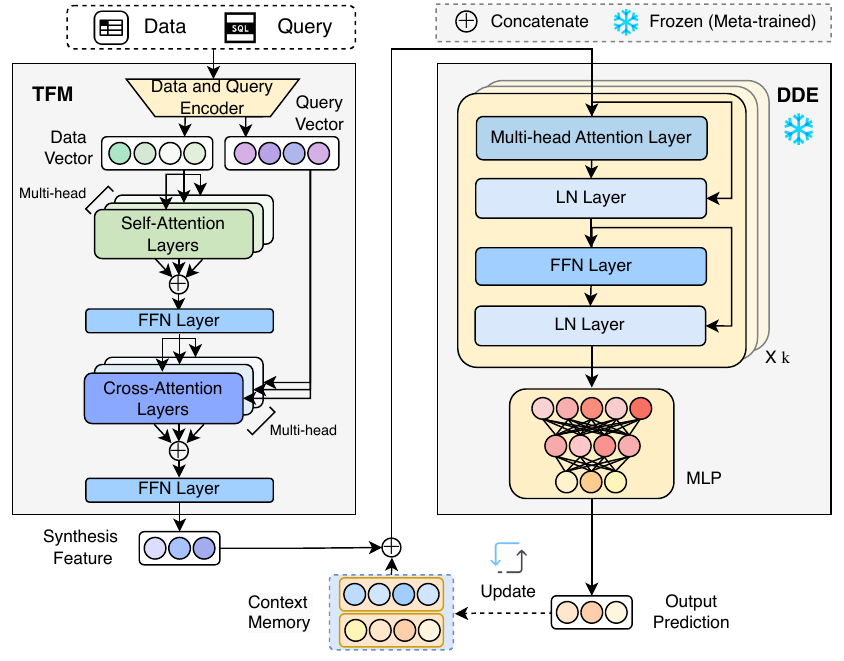}  \vspace{-2mm}
    \caption{The architecture of \name. }
    \label{fig:structure}
    \vspace{-4mm}
\end{figure}

\subsection{Dynamic Decision Engine} 
\label{sec:decision_engine}
The DDE forms the core module of \name.
As illustrated in Figure~\ref{fig:structure}, the DDE takes the task vector prepared by the TFM to provide real-time, context-aware predictions across various tasks.
It comprises two phases: \textit{Bayesian meta-training} and \textit{in-context adaptation}.
\ignore{
The DDE is meta-trained using synthetic datasets generated from curated priors, which is to pre-adapt it to evolving concepts in databases on a per-query basis.
In this subsection, we first introduce the novel Bayesian meta-training for DDE, and then, elaborate on how DDE supports in-context adaptation to concept drifts on-the-fly in databases, which maintains high prediction accuracy without the need for extra data collection and further explicit training.
}

\subsubsection{Bayesian Meta-training}
DDE is pre-trained using synthetic datasets sampled from prior distributions, which equips the model with broad generalization capabilities, enabling rapid adaptation to unseen tasks.
The meta-training is based on Bayesian inference theory.
Formally, for a given sample $\mathbf{x}$ with the evolving concept represented by a set of $c$ observed sample pairs $\mathcal{C}=\{(\mathbf{y}_i, \mathbf{x}_i)\}_{i=1}^{c}$ from the current task, the Posterior Predictive Distribution (PPD) of task predictive modeling is:
\begin{align}
p(\mathbf{y}|\mathbf{x}, \mathcal{C}) &= \int_{\Phi}{p(\mathbf{y}|\mathbf{x},\phi)p(\phi|\mathcal{C})}d\phi \\
&\propto \int_{\Phi}{p(\mathbf{y}|\mathbf{x},\phi)p(\mathcal{C}|\phi)p(\phi)}d\phi 
\label{eq:bayes}
\end{align}

\vspace{-3mm}
\noindent
where the task distribution $p(\phi)$ is sampled from curated prior distributions $\Phi$ to diversify the adaptability of DDE to different prediction tasks.
Notably, to capture complex dependencies and uncover underlying causal mechanisms,
we employ Bayesian Neural Networks (BNNs)~\cite{neal2012bayesian,gal2016uncertainty} and Structural Causal Models (SCMs)~\cite{pearl2009causality,peters2017elements} in constructing the prior distribution following PFNs~\cite{hollmann2022tabpfn}.

Based on the PPD formulation in Eq.~(\ref{eq:bayes}),
we first generate synthetic datasets, namely the concept $\mathcal{C}$ of observed samples from the task distribution $p(\phi)$, i.e., $\mathcal{C} \sim p(\mathcal{C}|\phi)$.
Second, we sample the data points $(\mathbf{x},\mathbf{y})$ for predictive modeling from $p(\mathbf{x},\mathbf{y}|\phi)$.
Next, we can train DDE using the input-output configuration via the loss:
\begin{align}
\mathcal{L}_{DDE}=\mathbb{E}_{( (\mathbf{x}, \mathcal{C}),\mathbf{y} )\in p(\phi)}[-\log q_{\theta}(\mathbf{y}|\mathbf{x},\mathcal{C})]
\label{eq:DDE_loss}
\end{align}

\vspace{-2mm}
\noindent
where the $q_{\theta}(\mathbf{y}|\mathbf{x},\mathcal{C})$ is the model's predictive distribution parameterized by $\theta$.
By minimizing this expected negative log probability $\mathcal{L}_{DDE}$, DDE is trained to maximize the likelihood of the observed data under the current task distribution $p(\phi)$.
In particular, $\mathcal{L}_{DDE}$ can be formalized as follows for different types of tasks, corresponding to regression and classification tasks, respectively.
\begin{small}
\begin{align}
\mathcal{L}_{reg}&= \mathbb{E}_{( (\mathbf{x}, \mathcal{C}),\mathbf{y} ) \in p(\phi)} \left[ \frac{(\mathbf{y} - \mu)^2}{2\sigma^2} + \log \sigma \right]\\
\mathcal{L}_{cls}&=\mathbb{E}_{((\mathbf{x},\mathcal{C}),\mathbf{y})\in p(\phi)}\left[-\sum_{k=1}^{K}\mathbb{I}_{\mathbf{y}=k}\log q_{\theta}(\mathbf{y}=k|\mathbf{x},\mathcal{C})\right]
\end{align}
\end{small}
where $\mu$ and $\sigma$
are the mean and standard deviation in regression tasks, $\mathbb{I}(\cdot)$ is the indicator function and $q_{\theta}(\mathbf{y}=k|\mathbf{x},\mathcal{C})$ is the predicted probability of class $k$ in classification tasks.

\vspace{-1mm}
\noindent \textbf{Remark.} We note that the Bayesian meta-training is performed only once on the curated prior distributions
across various tasks.
With Bayesian meta-training, \name
is enabled to quickly adapt to new concepts using a limited set of observed samples of the concept.
This offers several advantages:
(1) \textit{Cost-effective Data Collection}:
Generating synthetic data is significantly more cost-effective and faster than traditional data collection.
(2) \textit{One-time Effort}: The process is a one-time effort, eliminating frequent retraining after deployment. 
(3) \textit{No Privacy Issues}: Synthetic data does not contain real user information, thereby circumventing privacy and security concerns. 
\blue{(4) \textit{Scalability}: This strategy allows for easy adoption of desired prior task distributions instead of rebuilding the entire model from scratch.}


\subsubsection{In-context Adaptation}
During inference, we query the meta-trained DDE with the tuple $(\mathbf{z}_\mathcal{T},\mathcal{C})$ as input, where $\mathcal{C}=(\mathcal{Q}_{pmt},\mathcal{Y}_{pmt})$, termed as \textit{context memory}, contains contextual information of the current task.
$\mathcal{Q}_{pmt}$ and $\mathcal{Y}_{pmt}$ denote the sequences of recent queries and the system feedback, namely true outputs, which are organized into two separate first-in, first-out (FIFO) queues of size $\varrho$.
This strategy enables DDE to dynamically adapt to new concepts guided by the context memory during inference, thus avoiding backpropagation-based adaptation such as fine-tuning or retraining.


\vspace{-1mm}
\noindent \textbf{Remark.} To better understand the in-context adaptation mechanism, we examine the key differences between \name and existing learned approaches.
Existing methods like~\citet{marcus2021bao,zhao2022lightweight,wang2023speeding}
typically learn a static mapping from input to output as in Eq.~\ref{eq:fixed_map}, which assumes a fixed data distribution.
When concept drift occurs in the time interval $\Delta t=t'-t$, i.e., $\mathcal{D}_t \ne \mathcal{D}_{t'}$ and  $P_t \left(\mathbf{x},\mathbf{y} \right) \ne P_{t'}\left( \mathbf{x}, \mathbf{y} \right)$, the mapping $f_{\mathcal{D}_t,\Theta_t}$ from the input to the output should change accordingly.
To handle concept drift, these methods require collecting sufficient samples from the new distribution and updating the mapping $f_{\mathcal{D}_t,\Theta_t}$ with parameter $\Theta_t$ based on these samples, so as to obtain a new mapping function $f_{\mathcal{D}_{t'},\Theta_{t'}}$ with parameter $\Theta_t'$ that aligns with the new distribution $\mathcal{D}_{t'}$.
In contrast, our new paradigm essentially learns a \textit{conditional mapping} as formulated in Eq.~\ref{eq:dyn_map}, which explicitly models the evolving concept provided by the context memory $\mathcal{C}_t$ as the context of the current distribution $\mathcal{D}_{t}$.
\begin{align}
\label{eq:fixed_map}
\forall \,t,\;\; &f_{\mathcal{D}_t,\Theta_t}: \mathbf{x}\to \mathbf{y}\\
\forall \,t,\;\; &f_{\mathcal{D}_t,\Theta}:(\mathbf{x} \,| \,\mathcal{C}_t) \to \mathbf{y}
\label{eq:dyn_map}
\end{align}

\vspace{-3mm}
This adaptability via the in-context adaptation mechanism is well-suited for databases.
When a query is executed, the corresponding system output becomes immediately available and can be stored in the context memory to provide supervision for contextualized predictions of subsequent queries.
Also, For user-oriented tasks like data classification, the context memory within \name allows for online user feedback,  which facilitates the development of a customized system better aligned with user preferences.
\vspace{-1mm}
\subsection{\name Workflow: Training to Inference}\label{sec:Workflow}
\ignore{
\name comprises two cascading modules, the TFM module $\mathcal{M}_{TFM}$ parameterized by $\Theta_\mathcal{T}$ and the DDE module $\mathcal{M}_{DDE}$ parameterized by $\Theta_\mathcal{D}$, represented as follows:
\begin{align}
\label{eq:model_workflow}
\mathcal{M}( \mathbf{q}; \Theta_\mathcal{T},\Theta_\mathcal{D})=\mathcal{M}_{DDE}( \mathcal{M}_{TFM}(\mathbf{q}; \Theta_\mathcal{T}); \Theta_\mathcal{D})
\end{align}
Next, we elaborate on the workflow of \name, covering phases of offline training and online inference and adaptation.
}

\vspace{-1mm}
\noindent \textbf{Training.} \name is trained in two stages, as outlined in Algorithm~\ref{alg:training}.
(i) First, the $\mathcal{M}_{DDE}$ module undergoes a \textit{one-off meta-training} phase using $\mathcal{L}_{DDE}$ in Eq.~\ref{eq:DDE_loss} across crafted task distributions.
Note that the meta-training is not to optimize \name directly on end tasks but to prepare DDE to adapt to new tasks met during inference without further training.
(ii) Second, the $\mathcal{M}_{TFM}$ module is trained to extract informative latent features that are critical for the specific tasks at hand.
The training of TFM is tailored to optimize performance on these tasks.
This employs a task-specific loss $\mathcal{L}_{TS}$ to extract informative features for the DDE module.
%
%

\begin{algorithm}[hb]
\small
\caption{\name Training}
\label{alg:training} 
\begin{algorithmic}[1]
\renewcommand{\algorithmicrequire}{\textbf{Input:}}  
\renewcommand{\algorithmicensure}{\textbf{Output:}} 
\REQUIRE Designed priors $p(\phi)$, number of synthetic datasets $\mathcal{H}$, each with $N_o$ observed samples, queue size $\varrho$ in the context memory, learning rate $\eta_\mathcal{T}$ for $\mathcal{M}_{TFM}$ and $\eta_\mathcal{D}$ for $\mathcal{M}_{DDE}$.
\ENSURE \name $\mathcal{M}_F( \mathbf{x}; \Theta_\mathcal{T},\Theta_\mathcal{D})$ constructed by cascading $\mathcal{M}_{TFM}$ and $\mathcal{M}_{DDE}$ with parameters $\Theta_\mathcal{T}$ and $\Theta_\mathcal{D}$.

\STATE Initialize $\mathcal{M}_{TFM}$ and $\mathcal{M}_{DDE}$ with random weights $\Theta_\mathcal{T}$ and $\Theta_\mathcal{D}$

\FOR{$i=1$ {\bfseries to} $\mathcal{H}$}
     \STATE Sample synthetic datasets $\widetilde{D}_i \sim p(\mathcal{C}|\phi)$
     \STATE Randomly select context $\mathcal{C}$ based on $\{(\mathbf{x}_j,\mathbf{y}_j)\}_{j=1}^{\varrho}$ from $\widetilde{D}_i$
    \REPEAT
    \STATE Randomly select a training batch $\{(\mathbf{x}_j,\mathbf{y}_j)\}_{j=1}^{N_o}$ from $\widetilde{D}_i$
    \STATE Compute stochastic loss $\mathcal{L}_{DDE}$ using Eq.~\ref{eq:DDE_loss}
    \STATE Update $\Theta_\mathcal{D}$ using stochastic gradient descent $\Theta_\mathcal{D} \leftarrow \Theta_\mathcal{D} - \eta_\mathcal{D} \triangledown_{\Theta_\mathcal{D}}\mathcal{L}_{DDE}$
    \UNTIL{Convergence}
    
\ENDFOR
    \REPEAT
    \STATE Randomly sample a minibatch
    \STATE Update $\Theta_\mathcal{T}$ by minimizing the loss $\mathcal{L}_{TS}$ of the specific task $\Theta_\mathcal{T} \leftarrow \Theta_\mathcal{T} - \eta_\mathcal{T} \triangledown_{\Theta_\mathcal{T}}\mathcal{L}_{TS}$
    \UNTIL{Convergence}
 
 \STATE $\mathcal{M}_F(\mathbf{x}; \Theta_\mathcal{T}, \Theta_\mathcal{D})=\mathcal{M}_{DDE}( \mathcal{M}_{TFM}(\mathbf{x}; \Theta_\mathcal{T}); \Theta_\mathcal{D})$;

\STATE {\bfseries Return} \name $\mathcal{M}_F$
\end{algorithmic}
\end{algorithm}

\vspace{-1mm}
\noindent \textbf{Inference.} Once trained, \name is ready for concurrent online inference and adaptation
in a real-time environment:
\vspace{-1mm}
\ignore{
Specifically, for an incoming input query $\mathbf{q}$, the TFM first extracts its task vector as shown in Eq.~\ref{eq:inference_p1} below.
The task vector, along with its contextual information in context memory $\mathcal{C}=(\mathbf{Q_{pmt}},\mathbf{Y_{pmt}})$, 
are then fed to the DDE module as Eq.~\ref{eq:inference_p2}. 
After executing the current query, the query and the corresponding ground-truth result returned by the database are used to update the context memory $\mathbf{Q_{pmt}}$ and $\mathbf{Y_{pmt}}$.}
\begin{align}
&\scalebox{0.91}{$\mathbf{x} \Rightarrow \mathcal{M}_{TFM}(\mathbf{x}; \Theta_\mathcal{T})=\mathbf{z}_\mathcal{T} \Rightarrow  \mathcal{M}_{DDE}(\mathbf{z}_\mathcal{T},\mathcal{C}; \Theta_\mathcal{D})=\mathbf{y}$}\\
&\scalebox{0.95}{$\mathbf{x}\Rightarrow\mathcal{S}_{execute}(\mathbf{x})=\mathbf{y}^{*} \Rightarrow (\mathbf{z}_\mathcal{T},\mathbf{y}^{*})\xrightarrow{\text{update}} \mathcal{C}$}
\end{align}

\vspace{-3mm}
where $\mathcal{S}_{execute}(\cdot)$ is the data system executor that produces the actual system output $\mathbf{y}^{*}$.
The process is detailed in Algorithm~\ref{alg:inference}.
Fundamentally, \name streamlines the model update process by replacing the traditional, cumbersome backpropagation with an efficient forward pass via meta-training and an in-context adaptation mechanism.

\begin{algorithm}[t]
\small
\caption{Concurrent \name Inference and Adaptation}
\label{alg:inference} 
\begin{algorithmic}[1]
\renewcommand{\algorithmicrequire}{\textbf{Input:}}  
\renewcommand{\algorithmicensure}{\textbf{Output:}}  
\REQUIRE $\mathcal{M}_{TFM}$ and $\mathcal{M}_{DDE}$ with parameters $\Theta_\mathcal{T}$ and $\Theta_\mathcal{D}$, input query and data underlying the data system.
\ENSURE Predicted output $\mathbf{y}$.

\STATE Extract latent feature $\mathbf{z}_\mathcal{T}$ incorporating information from query and data, using $\mathcal{M}_{TFM}$ as $\mathbf{z}_\mathcal{T}= \mathcal{M}_{TFM}(\mathbf{x}; \Theta_\mathcal{T})$
\STATE Gather context memory
$\mathcal{C}=(\mathcal{Q}_{pmt},\mathcal{Y}_{pmt})$
\STATE Predict $\mathbf{y}$ by inputting latent feature $\mathbf{z}_\mathcal{T}$ and context memory $\mathcal{C}$ into $\mathcal{M}_{DDE}$ as $\mathbf{y}= \mathcal{M}_{DDE}(\mathbf{z}_\mathcal{T},\mathcal{C}; \Theta_\mathcal{D})$
\STATE Store $\mathbf{z}_\mathcal{T}$ and the corresponding system output $\mathbf{y}^{*}$ into queue $\mathcal{Q}_{pmt}$ and $\mathcal{Y}_{pmt}$ to update the context memory $\mathcal{C}$
\STATE Remove oldest entries from $\mathcal{Q}_{pmt}$, $\mathcal{Y}_{pmt}$ to maintain size $\varrho$
    
\STATE {\bfseries Return} $\mathbf{y}$
\end{algorithmic}
\end{algorithm}

\icmlblue{
\name efficiently accommodates large dynamic databases through incremental histogram maintenance in $O(N_v)$ with $N_v$ modified records and adapts to concept drift using a FIFO key-value memory for in-context adaptation.
The cross-attention mechanism operates on a single query vector and incurs only a linear overhead of $O(d_a\varrho)$, where $d_a$ is the attention dimension in DDE.
}
This flexible and scalable workflow ensures that \name learns effectively from new tasks on-the-fly, adapting to evolving concepts in dynamic databases.
\icmlblue{
\vspace{-1mm}
\subsection{Model Generalization Error Bound Analysis}}
\vspace{-1mm}
In this section, we analyze the generalization error bounds of \name against conventional models optimized for
static data,
when faced with post-training data evolving.
We aim to uncover the susceptibility of outdated static models to dynamic environments and showcase \name's resilience.\\
Consider a model $\hat{f}_i$ trained on dataset $D^i$ and frozen once training concludes.
Subsequent $k$ single-point data operations alter the data from $D^i$ to $D^j$, where each operation is atomic, comprising either insertion or deletion\footnote{For simplicity, we solely consider insertion and deletion since the update operation can be decomposed into these operations.}.
$f_{D^j}$ refers to
the ground-truth mapping to $D^j$.
We now explore the worst-case bound on expected maximum generalization error for robustness.

\vspace{-1mm}
\begin{theorem}[]
Consider a model $\hat{f}_i$ trained on an initial dataset $D^i$, where $|D^i|=i$. After $k$ data operations, including $s$ insertion and $r$ deletion, we obtain a new dataset $D^j$ of size $|D^j|=j$, where $k=s+r>1$ and the net difference in data size $|j-i|=|s-r|$. Suppose data in $D^j$ are i.i.d from any continuous distribution $\chi$, we have
\begin{align*}
\sup_{\mathbf{x}} \; 
\mathbb{E}_{D^j \sim \chi}
\bigl[
  \bigl|
   \hat{f}_i(\mathbf{x}) 
    \;-\; 
    f_{D^j}(\mathbf{x})
  \bigr|\bigr]
\;\ge\; k-1
\end{align*}
\label{theorem:static}
\vspace{-6mm}
\end{theorem}

Theorem~\ref{theorem:static} states that the risk of using a stale model to make predictions escalates at a minimum rate of $\Omega(k)$
as data evolves.
Theoretically, to sustain a
error at $\epsilon$, $\frac{\varkappa}{\epsilon+1}$ model retraining is needed for every $\varkappa$ data operation. 
The cost per retraining session generally involves processing the entire dataset or a significant portion thereof in the scale $\mathcal{O}(\varkappa)$~\cite{zeighami2024theoretical}. 
Consequently, the amortized cost per data operation, given that retraining the model every $\epsilon+1$ data operation, is also $\mathcal{O}(\varkappa)$.
Thus, maintaining low error rates in such a dynamic setting can be computationally expensive.
In contrast, our model defined as $\hat{f}(\mathbf{x}| \mathcal{C}^j)$ exhibits resilience to changes in data. 
\begin{theorem}[]
Consider \name trained when the underlying database is $D^i$ and using context memory $\mathcal{C}^j$ to perform prediction when the database evolves to $D^j$, we have
\begin{align*}
\sup_{\mathbf{x}} \mathbb{E}_{D^j\sim \chi}\Bigl[ \bigl| \hat{f}(\mathbf{x}|  \mathcal{C}^j) - f_{D^j}(\mathbf{x}) \bigr| \Bigr] \leq \frac{\aleph}{\sqrt{\varrho}}
\end{align*}
with high probability $1-\delta$, where $\aleph =\sqrt{\frac{1}{2}(\kappa +\ln{\frac{1}{\delta}})} +\sqrt{\frac{\pi}{2} }$.
Here, $\varrho$ is the size of the context memory $\mathcal{C}^j$, $\kappa$ is a constant reflecting the training adequacy,
and data in $D^j$ is drawn i.i.d from any continuous distribution $\chi$.
\label{theorem:FLAIR}
\vspace{-2mm}
\end{theorem}
Theorem~\ref{theorem:FLAIR} demonstrates that the generalization error of \name can be effectively controlled by the size of context memory $\varrho$.
By ensuring that $\varrho$ is sufficiently large, the generalization error remains well within the bounds of $\mathcal{O}(\frac{1}{\sqrt{\varrho}})$.
Unlike traditional models that experience a linear growth in generalization error with each data operation $k$, \name's error remains stable regardless of $k$, showing no performance deterioration with post-training data changes.
Specifically, setting $\varrho$ to be at least $(\frac{\aleph}{k-1})^2$ ensures that the expected worst-case generalization error of \name stays below static models.
This aligns with existing research~\cite{namkoong2016stochastic,sagawadistributionally} that considers potential distribution shifts during training bolsters model resilience after deployment.
Overall, Theorem~\ref{theorem:FLAIR} elucidates \name's theoretical superiority over static models in maintaining continuous accuracy and operational efficiency, providing a scalable solution with frequent data evolving.


\section{Experiments}
\label{sec:experiment}
In this section, we systematically evaluate the effectiveness, efficiency, and transferability of \name.
\icmlblue{
Extensive experiments are conducted on real-world benchmarks for cardinality estimation to test the effectiveness of \name across various degrees of concept drift, followed by assessments of training and inference efficiency.
We then explore \name's robustness against long-term
concept drift, and its transferability to representative user-oriented tasks within databases.
Moreover, we integrate \name with PostgreSQL to confirm its compatibility with operational environments.}


\begin{table*}[t]
\centering
\renewcommand{\arraystretch}{1}
\caption{Overall performance of cardinality estimation task under concept drift. The best performances are highlighted in bold and underlined, and the second-best are bold only.}
\vspace{-2mm}
\resizebox{2.1\columnwidth}{!}{
\begin{tabular}{llcccccccccccccccccc}
\toprule
\multirow{3}{*}{Data} & \multirow{3}{*}{Method} & \multicolumn{9}{c}{Mild Drift} & \multicolumn{9}{c}{Severe Drift} \\
\cmidrule(lr){3-11} \cmidrule(lr){12-20}
& & \multirow{2}{*}{GMQ} & \multicolumn{4}{c}{Q-error} & \multicolumn{4}{c}{P-error} & \multirow{2}{*}{GMQ} & \multicolumn{4}{c}{Q-error} & \multicolumn{4}{c}{P-error} \\
& & & 50\% & 75\% & 90\% & 95\% & 50\% & 75\% & 90\% & 95\% & & 50\% & 75\% & 90\% & 95\% & 50\% & 75\% & 90\% & 95\% \\
\midrule

\multirow{5}{*}{STATS} 
& Fine-tune & \textbf{5.35} & 3.47 & 9.32 & 33.99 & 77.93 & 6.21 & 17.72 & 54.21 & 111.85 & \textbf{\underline{5.02}} & 3.35 & 6.44 & 17.44 & 65.76 & 10.24 & 55.92 & 255.82 & 927.08 \\
& PostgreSQL & 174.38 & 497.56 & 611.53 & 21556.35 & 70977.46 & 8.87 & 52.29 & 157.93 & 174.24 & 293.47 & 758.89 & 6740.46 & 62020.12 & 218196.66 & 10.39 & 83.37 & 401.75 & 1296.15 \\
& ALECE & 20.29 & 15.03 & 52.26 & 197.61 & 430.69 & 7.67 & 30.05 & 131.25 & 249.24 & 36.16 & 22.77 & 112.79 & 624.31 & 1172.69 & 8.61 & 48.72 & 312.12 & 1130.76 \\
& DDUp & 5.79 & 4.49 & 10.20 & 26.41 & 72.68 & 8.00 & 29.59 & 64.91 & 241.27 & 10.95 & 9.51 & 20.22 & 46.15 & 87.18 & 13.61 & 43.92 & 109.24 & 216.64 \\
& FLAIR & \textbf{\underline{4.49}} & 2.86 & 6.93 & 24.06 & 60.94 & 7.01 & 28.04 & 61.70 & 162.61 & \textbf{5.47} & 3.12 & 7.87 & 28.52 & 81.57 & 7.97 & 26.78 & 308.43 & 1005.64 \\

\midrule

\multirow{6}{*}{Job-light}
& Fine-tune & \textbf{2.45} & 1.36 & 2.05 & 9.28 & 20.47 & 1.09 & 2.56 & 3.42 & 4.09 & \textbf{8.09} & 2.31 & 9.68 & 57.51 & 5168.29 & 1.02 & 1.08 & 1.74 & 1.86 \\
& PostgreSQL & 9.36 & 1.89 & 6.93 & 21.42 & 87.12 & 1.28 & 2.14 & 3.98 & 7.06 & 32.09 & 11.75 & 282.67 & 3834.32 & 7200.49 & 1.90 & 2.78 & 4.20 & 62.43 \\
& DeepDB & 32.28 & 10.52 & 436.77 & 698.12 & 6894.09 & 1.98 & 18.19 & 36.24 & 126.55 & 49.69 & 14.76 & 972.51 & 7864.98 & 7.65e\textsuperscript{5} & 1.77 & 17.31 & 22.31 & 51.89 \\
& ALECE & 12.21 & 11.59 & 19.34 & 26.40 & 63.37 & 1.96 & 4.83 & 8.72 & 19.06 & 27.32 & 11.72 & 114.32 & 1920.34 & 6970.01 & 1.56 & 2.35 & 3.68 & 4.26 \\
& DDUp & 4.16 & 3.60 & 4.99 & 15.62 & 46.98 & 1.59 & 2.14 & 3.88 & 5.12 & 10.96 & 6.65 & 35.35 & 162.51 & 203.34 & 1.09 & 1.65 & 1.89 & 2.79 \\
& FLAIR & \textbf{\underline{2.36}} & 1.29 & 2.28 & 6.93 & 18.62 & 1.18 & 1.36 & 2.94 & 4.26 & \textbf{\underline{7.95}} & 2.38 & 10.21 & 29.71 & 4826.64 & 1.03 & 1.41 & 1.78 & 2.38 \\
\bottomrule
\end{tabular}
\label{tab:ce_effective}
}
\vspace{-4mm}
\end{table*}

\vspace{-2mm}
\icmlblue{
\subsection{Experimental Setup}}
\label{sec:exp_setup}

\vspace{-1mm}
\noindent \textbf{Benchmarks.}
We evaluate \name on two established real-world benchmarks:
\textbf{STATS}~\cite{STATS-web} and \textbf{JOB-light}~\cite{leis2018query,leis2015good}.
STATS contains over 1 million records,
while JOB-light, derived from the IMDB dataset, includes 62 million records.
\ignore{
As in a recent work~\cite{li2023alece},
our evaluation involves randomly generating 2000 varied queries for each benchmark for training and using realistic queries as test sets.}
We simulate real-world database conditions in our experiments by incorporating varied SQL operations
and design scenarios that mirror different levels of concept drift, ranging from mild to severe.

\vspace{-4mm}
\begin{itemize}[leftmargin=*]
\setlength\itemsep{0mm}
\item STATS~\cite{STATS-web}, includes 8 relations with 43 attributes.
There are 1,029,842 records from the anonymized Stats Stack Exchange network.
The benchmark workload includes 146 queries, featuring both PK-FK and FK-FK join. 

\item JOB-light~\cite{leis2018query,leis2015good}, derives from a subset of the IMDB dataset and encompasses 6 relations with 14 attributes. 
There are 62,118,470 records in total.
The benchmark workload consists of 70 queries focusing on the PK-FK join. 
\vspace{-4mm}
\end{itemize}

As in a recent work~\cite{li2023alece}, 
we randomly generate 2000 diverse queries with sub-queries to form the training set for each benchmark.
In the STATS benchmark, we utilize an existing workload of 146 queries with 2603 sub-queries as the test set. 
For JOB-light, the test set comprises 70 queries associated with 696 sub-queries.
Additionally, we incorporate a dynamic workload into each benchmark's training and test sets. 
This dynamic workload includes a variety of SQL operations, including insert, delete, and update, strategically varied in proportion throughout different phases of the experiment.
Notably, the ground truth for the queries is obtained by executing them, as both the dynamic workload and data changes can influence the results over time. 
For the CE task, queries yielding a ground-truth cardinality of zero are excluded from the analysis to ensure data integrity and relevance.

\vspace{-1mm}
\noindent \textbf{Downstream Tasks.}\label{sec:applications}
We primarily assess \name's core performance
through cardinality estimation (CE) tasks, alongside exploring its capabilities in user-oriented activities like approximate query processing (AQP) and in-database data analytics involving data classification and regression.

\vspace{-4mm}
\begin{itemize}[leftmargin=*]
\setlength\itemsep{0mm}
\item Cardinality Estimation (CE) estimates the number of rows a query returns, aiding query planners in optimizing execution plans.

\item Approximate Query Processing (AQP) quickly delivers approximate
results from large datasets by balancing accuracy with computational efficiency.

\item In-database Data Analytics involves data classification and regression tasks executed within the database engine, delivering insights directly from the data source.
(i) Data classification boosts business intelligence by using categorical attributes to categorize tuples, such as product types and transaction statuses, supporting 
analytics in database systems.
(ii) Data regression predicts continuous outcomes, enhancing predictive analytics and decision-making on platforms like Oracle~\cite{helskyaho2021oracle} and Microsoft SQL Server~\cite{maclennan2011data,harinath2008testing}.
\end{itemize}


\vspace{-1mm}
\noindent \textbf{Baselines.}
\label{sec:baseline}
We compare \name with predominant families of CE technologies, including the estimator from 
PostgreSQL~\cite{post1996}, and SOTA learned approaches for
dynamic environments, such as DeepDB~\cite{hilprecht2019deepdb}, ALECE~\cite{li2023alece}, and DDUp~\cite{kurmanji2023detect} with NeuroCard~\cite{yang2020neurocard} being used as its base model.
We also compare \name with model fine-tuning outlined in~\cite{kurmanji2023detect}, serving as a high-performance baseline despite being computationally intensive.
For AQP, our baselines include DBest++~\cite{ma2021learned}, which utilizes only frequency tables (FTs) for the update, DBest++FT, which updates both FTs and mixture density networks (MDNs), and DDUp, which uses DBest++ as its base model.
For in-database data analytics, we compare \name with AutoML system AutoGluon~\cite{erickson2020autogluon} and established ML algorithms, including K-nearest-neighbors (KNN), RandomForest, MLP, and popular
boosting methods, XGBoost~\cite{chen2016xgboost}, LightGBM~\cite{ke2017lightgbm} and CatBoost~\cite{prokhorenkova2018catboost} for data classification, and AutoGluon, SVR, MLP, DecisionTree, RandomForest, and GradientBoosting for regression.

\noindent \textbf{Implementation.}
\blue{\name is implemented in Python with Pytorch 2.0.1.
The baseline methods are implemented using open-source packages or source code provided by the original researchers, adhering to recommended settings.
The experiments involving PostgreSQL are conducted on PostgreSQL 13.1.}
All the experiments are conducted on a server with a Xeon(R) Silver 4214R CPU @ 2.40GHz (12 cores), 128G memory, and a GeForce RTX 3090 with CUDA 11.8.
The OS is Ubuntu 20.04 with Linux kernel 5.4.0-72.

\vspace{-1mm}
\noindent \textbf{Evaluation Metrics.}
We evaluate \name's effectiveness and efficiency across various tasks using targeted metrics.
(1) Effectiveness Metrics:
For CE tasks, we report the accuracy by the geometric mean of the Q-error (GMQ) as~\cite{li2022warper,dutt2019selectivity} along with Q-error and P-error across various quantiles, with particular emphasis on the tail performance.
For AQP tasks, we use mean relative error (MRE) to evaluate the accuracy of query approximations.
Additionally, we apply accuracy and F1 score for data classification and mean squared error (MSE) and the coefficient of determination ($R^2$) for data regression.
(2) Efficiency Metrics:
We assess \name's efficiency by examining storage overhead, building time, inference time, and adaptation time.

\begin{figure}[t]
    \centering
 \includegraphics[width=1\linewidth]{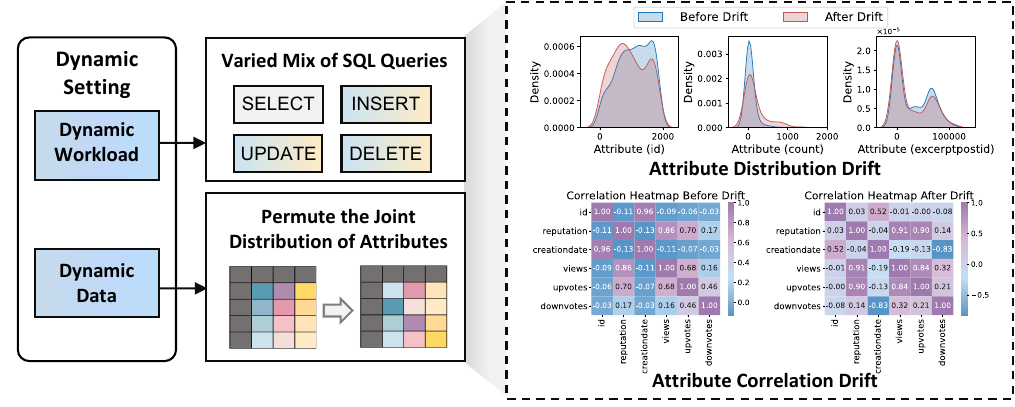} \vspace{-5mm}
    \caption{Overview of dynamic settings, illustrated by distribution discrepancies confirmed by Kolmogorov-Smirnov test p-values below 0.01 pre- and post-concept drift. 
    }
    \label{fig:dynamic_settings}
    \vspace{-4mm}
\end{figure}

\ignore{
\noindent \textbf{Efficiency Metrics.}
We evaluate the efficiency of \name within PostgreSQL by testing its query execution \underline{\textit{latency}}, which directly connects to the query optimizer and objectively shows how our method can enhance DBMS query performance. 
\underline{\textit{storage overhead}} gauges the memory requirement of a method, while \underline{\textit{building time}} measures the necessary offline training duration. 
\underline{\textit{inference time}} indicates the average time per input instance for estimation, crucial for real-time application. 
Lastly, \underline{\textit{adaptation time}} reflects how quickly the model can adjust to concept drift.
}

\noindent \textbf{Dynamic Settings and Data Drift.}
In our study, we explore a dynamic data system marked by variations in both workload and data, which is illustrated in Figure~\ref{fig:dynamic_settings}.
To emulate a real system environment, we introduce significant data drift after training and before testing. 
This involves sorting each column to alter the joint distribution of attributes and then performing random sampling from this permuted dataset.
The impact of these manipulations on data distribution and attribute correlations is visually depicted through histograms and heat maps in Figure~\ref{fig:dynamic_settings}, showcasing the data characteristics before and after experiencing data drift.
This dynamic scenario comprehensively mirrors real-world database operations where frequent insert, delete, and update actions induce gradual changes in data distribution. 
Over time, these incremental modifications accumulate, resulting in more pronounced shifts in data structures and inter-attribute relationships.
To rigorously assess the robustness of our approach, we design two scenarios based on the extent and nature of the changes.
\vspace{-4mm}
\begin{itemize}[leftmargin=*]
\setlength\itemsep{0mm}
\item Mild Drift: 
We permute and sample the data with 50\% of the dataset experiencing drift, testing the model response to moderate yet significant changes in data without additional data manipulations.

\item Severe Drift: We escalate the challenge by not only permuting and sampling 60\% of the data but also integrating 10\% random data manipulations, including additions, deletions, and value replacements to assess model capability under severe data transformations.
\end{itemize}

\begin{figure}[t]
    \centering
\includegraphics[width=1\linewidth,height=30mm]{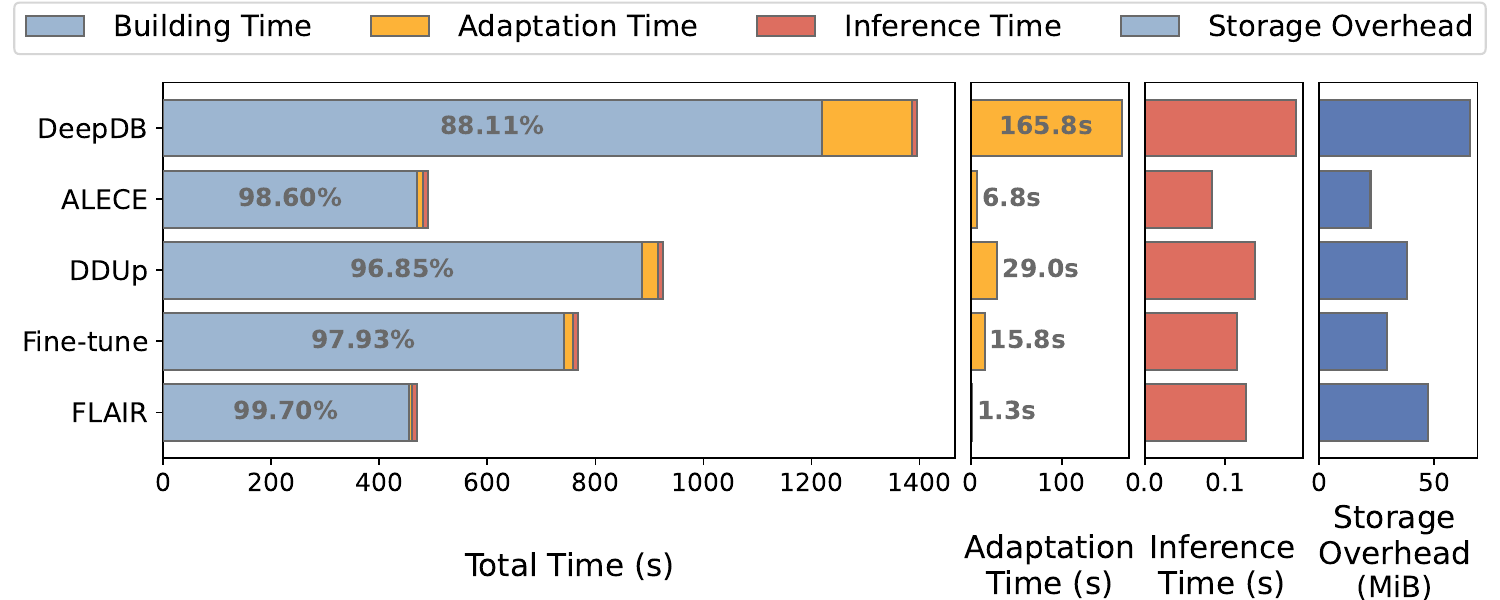} \vspace{-5mm}
    \caption{Comparison of model efficiency.
    }
    \label{fig:efficiency}
    \vspace{-4mm}
\end{figure}

\vspace{-1mm}
\subsection{Effectiveness}
\vspace{-1mm}
\label{sec:exp_effectiveness}
In~\cref{tab:ce_effective}, we report the overall performance comparison in CE task.
\ignore{
Data-driven approaches such as DeepDB and DDUp are configured with the database data after executing all statements from the training portion of the workload, reflecting a real-world system scenario as described in~\cite{li2023alece}.
DeepDB is not compared on STATS as it only supports PK-FK joins.
For \name, the queue size $\varrho$ is set to 80, unless specified otherwise. }
%
The results reveal that \name consistently delivers superior performance across all datasets and
dynamic scenarios, often matching or even surpassing the outcomes of the fine-tune approach.
Specifically, \name achieves the best performance in 29 out of 32 quantile metrics.
Even when including fine-tune comparisons, \name surpasses
nearly half of the evaluations for all metrics, underscoring its considerable
precision in dynamic environments.
Additionally, \name significantly outperforms PostgreSQL across all datasets and settings, highlighting the limitations of PostgreSQL's independence assumption that often results in inaccuracies with non-uniform data distributions. 
Furthermore, our experiments reveal that existing methods, including those using fine-tuning and knowledge distillation, struggle with rapid and complex changes in dynamic systems. In contrast, \name excels by promptly adapting to current concepts during concept drift, without data recollection, offline updates, or separate drift detection processes.

\ignore{
Additionally, \name significantly outperforms PostgreSQL on all datasets and settings.
This discrepancy highlights the limitations of the independence assumption commonly used in PostgreSQL, which often leads to inaccuracies in scenarios with non-uniform data distributions.
Furthermore, our experiments indicate that existing methods, even those that incorporate system dynamics through techniques such as fine-tuning and knowledge distillation, struggle to effectively address the complex and real-time changes occurring within dynamic systems. 
In contrast,
\name promptly adjusts based on contextual information, ensuring accurate predictions during concept drift without additional data collection, offline model updates, or separate drift detection processes.
}

\begin{figure}[t]
    \centering
    \includegraphics[width=1\linewidth, height=29mm]{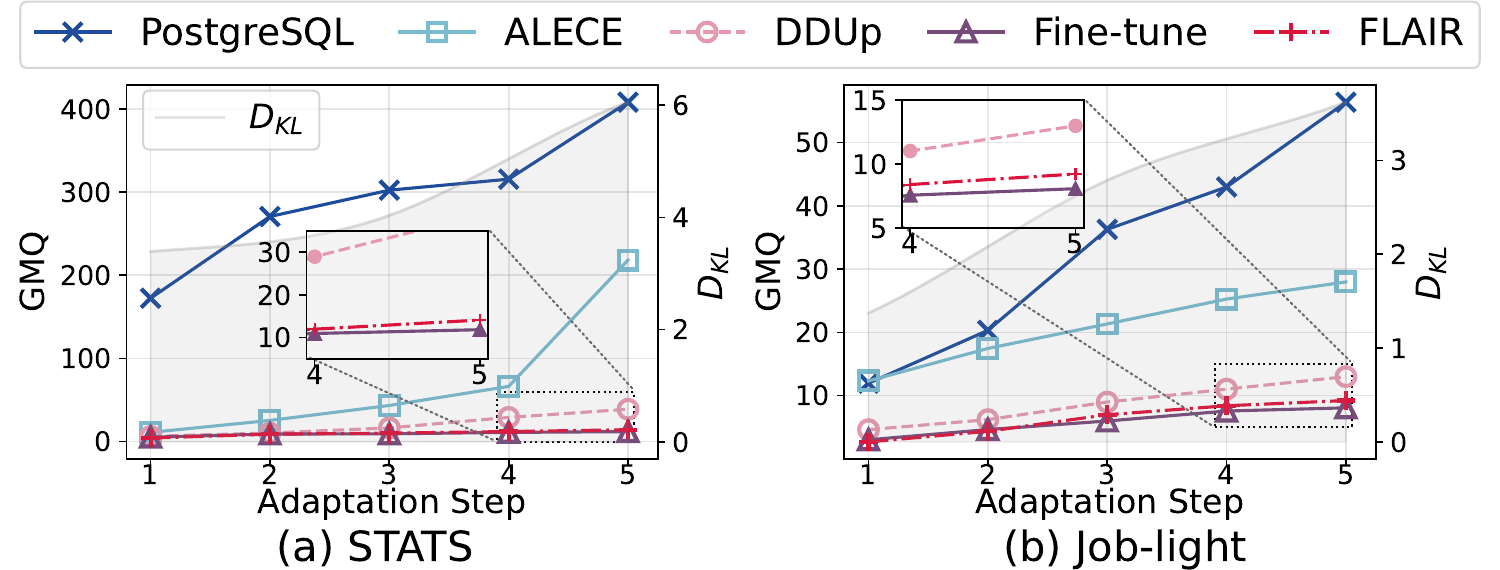} \vspace{-4mm}
    \caption{Comparison of model robustness for long-term incremental concept drift.
    }
    \label{fig:long_term}
     \vspace{-5mm}
\end{figure}

\subsection{Efficiency}
\label{sec:exp_efficiency}
We evaluate the construction efficiency and resource usage of \name alongside baseline models on the JOB-light benchmark. 
The results in Figure~\ref{fig:efficiency}
demonstrate that \name is notably efficient in both building and adaptation phases. 
Remarkably, \name accelerates adaptation speed by $5.2\times$ while reducing the GMQ by 22.5\% compared with the best baseline.
To further improve \name's inference efficiency, we implement an embedding caching mechanism in \name, which eliminates redundant computations by preventing recomputation on the repeated inputs. 
This enhancement significantly accelerates the inference process, yielding competitive inference times.
Taking the overall performance into consideration,
the slightly higher storage requirement imposed by \name is acceptable.

    

\vspace{-1mm}
\subsection{Long-term Incremental Concept Drift}\label{sec:long-term}
\vspace{-1mm}
To further assess \name's adaptability, we track the performance on STATS and JOB-light, focusing on gradual drift indicated by rising Kullback-Leibler divergence $D_{KL}$ over extended periods.
Figure~\ref{fig:long_term} illustrates that \name effectively handles the challenging conditions of long-term incremental concept drift across both benchmarks,
even on par with model fine-tuning.
Furthermore, we observe that
DDUp based on knowledge distillation is inferior to fine-tuning under long-term gradual drift. 
This is in line with the results in~\cref{sec:exp_effectiveness}, highlighting the inherent limitations of knowledge distillation: it mitigates catastrophic forgetting by preserving prior learned knowledge but can inadvertently replicate past errors, whereas fine-tuning directly adjusts to new data, correcting inaccuracies and adapting to evolving distributions.
%
Conversely, \name's innovative in-context adaptation paradigm, guided by dynamic context memory, achieves negligible error accumulation and ensures sustained adaptability without further training, distinguishing it from both knowledge distillation and fine-tuning.
%

\vspace{-1mm}
\subsection{Transferability}\label{sec:exp_generality}
\vspace{-1mm}
In data systems, system-internal tasks like CE provide immediate critical outcomes for optimization, while it is often not straightforward for user-oriented tasks.
Next, we validate \name's performance in user-oriented scenarios to showcase its wide applicability, where our context memory establishes a virtuous cycle of user feedback to refine model performance
and facilitate system customization.

\ignore{
In database systems, system-internal tasks such as cardinality estimation, database tuning, and transaction throughput measurement provide immediate ground-truth outcomes critical for query optimization, resource management, and system performance monitoring. 
However, obtaining ground truth for certain user-oriented tasks is more complex.
In these cases, our context memory in \name establishes a virtuous cycle of user feedback, where users can provide feedback on the model's predictive outcomes.
This feedback acts as a practical form of ground truth, facilitating continuous refinement of model performance on user-oriented tasks and enabling system customization.
Additionally, \name maintains dynamic context memory that blends user feedback data with recent model interactions, which will be further analyzed in~\cref{sec:appendix-ablation}.
Next, we will validate \name's performance in user-oriented scenarios, particularly in AQP and ML tasks, to showcase its wide applicability.
}

\begin{figure}[t]
    \centering
    \includegraphics[width=1\linewidth]{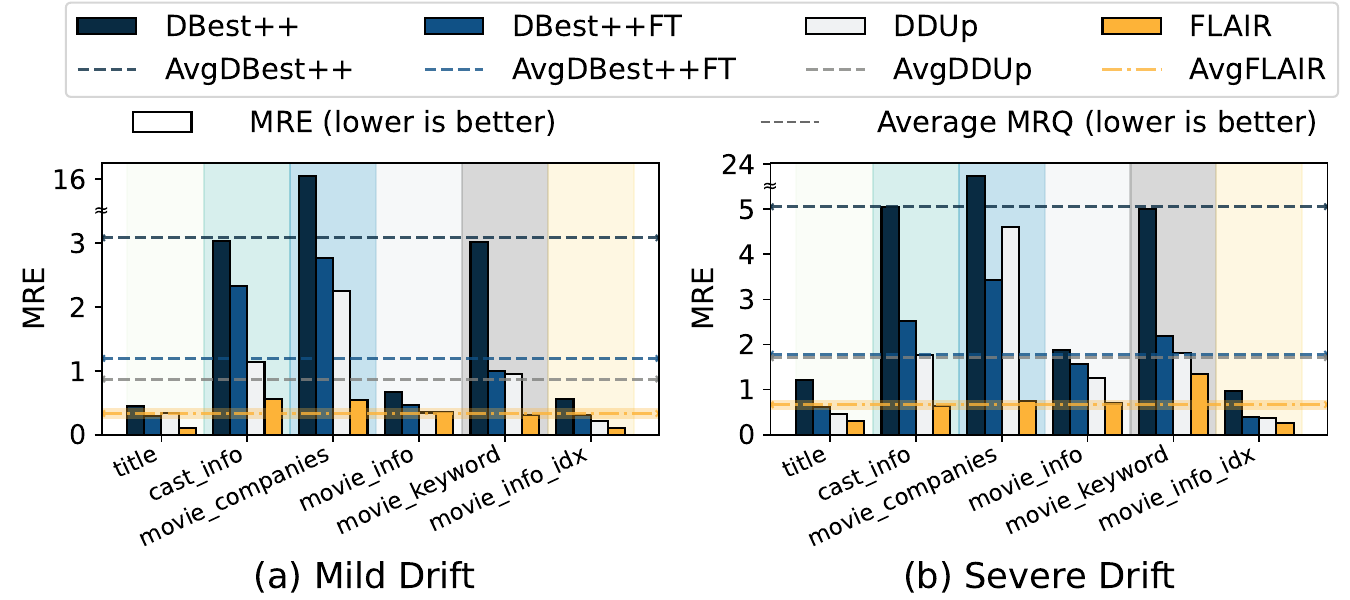} \vspace{-4mm}
    \caption{Performance of AQP task under concept drift.}
    \label{fig:aqp}
\end{figure}

\vspace{-1mm}
\textbf{Approximate Query Processing.}
The results in Figure~\ref{fig:aqp}, measured in MRE, consistently show that \name outperforms baseline approaches. 
Across various relations and dynamic settings, \name achieves significant error reductions, with averages up to or exceeding 10$\times$ with DBest++, 3$\times$ with DBest++FT, and 2$\times$ with DDUp. 
These findings highlight the effectiveness of \name in handling complex query scenarios.
Most of the time, \name outperforms methods that rely on fine-tuning and knowledge distillation, such as DBest++FT and DDUp.
This superiority stems from the limitations associated with only updating models during significant data drifts, which may not suffice for the accurate execution of AQP tasks in real and live system scenarios.
\ignore{
In practical deployments, offline model updates will lead to considerable system latency due to maintenance, or escalate errors when outdated models remain operational during updates.
Consistent with conclusions from prior research~\cite{kurmanji2023detect}, the knowledge distillation-based method generally outperforms fine-tuning, likely due to the use of lower learning rates that may hinder the model's ability to capture current data distributions.}

\begin{figure}[t]
    \centering
\includegraphics[width=1\linewidth]{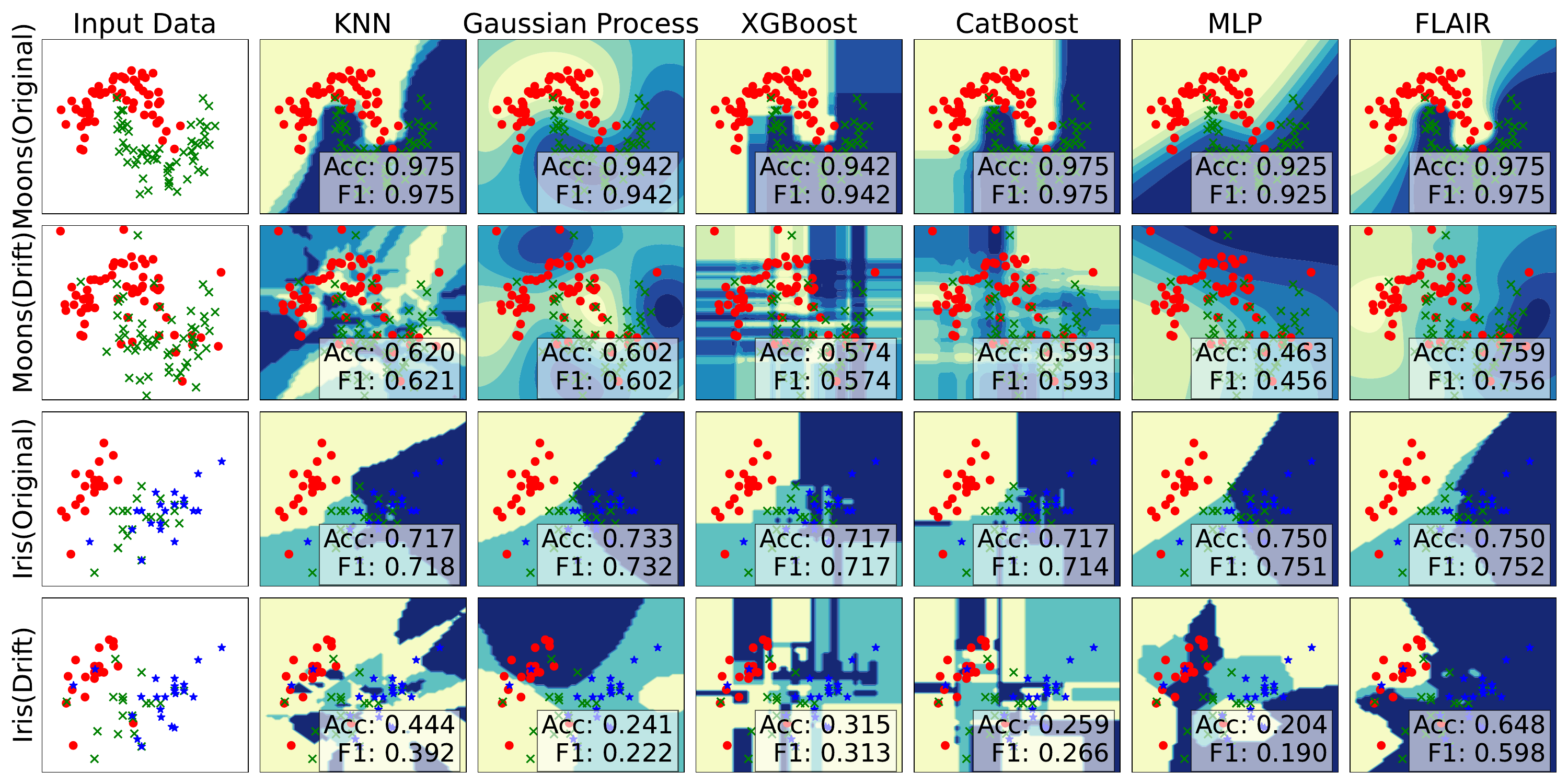} \vspace{-2mm}
    \caption{Decision boundaries and model performance on data classification task under concept drift.
    }
    \label{fig:toy_clf}
    \vspace{-2mm}
\end{figure}

\vspace{-1mm}
\textbf{In-database Data Analytics.}
We initially conduct a qualitative evaluation on illustrative toy problems to understand the behavior of \name under concept drift, comparing against standard classifiers as shown in Figure~\ref{fig:toy_clf}.
We utilize moons and iris datasets from scikit-learn~\cite{pedregosa2011scikit}.
For the drift scenarios,
we allocate 10\% of data for model updates and the remaining 90\% for evaluation. 
In each case, \name effectively captures the decision boundary between samples, delivering well-calibrated predictions.
We extend our empirical analysis to real-world
tasks, applying data classification for sentiment analysis and data regression for rating prediction on IMDB.
\ignore{
We compare \name with AutoML system AutoGluon~\cite{erickson2020autogluon} and established ML algorithms, including K-nearest-neighbors (KNN), RandomForest, MLP, and popular tree-based boosting methods, XGBoost~\cite{chen2016xgboost}, LightGBM~\cite{ke2017lightgbm} and CatBoost~\cite{prokhorenkova2018catboost} for data classification.
For regression, we include AutoGluon, SVR, MLP, DecisionTree, RandomForest, and GradientBoosting as our baselines.
}
\textbf{(i) Data Classification.}
We conduct sentiment analysis~\cite{maas2011learning} on IMDB, which is a prevalent binary classification task. 
We allocate 50\% of the original data as the training set, and following prior setups, induced data drift on the remaining data. We designate 20\% of the post-drift data as the update set and the remaining post-drift data as the test set.
For models that support incremental updates, such as XGBoost, LightGBM, CatBoost, and MLP, we incrementally update the models initially trained on the training set using the update set, while others are retrained on the update set.
Finally, we evaluate all models on the test set to measure their effectiveness in adapting to data drift, as summarized in Table~\ref{tab:clf}. 
The mean time represents the total execution time, integrating building, adaptation, and inference time averaged across two drift scenarios.
Our \name distinctly showcases its robustness and adaptability in handling concept drift, resulting in superior performance across both mild and severe drift scenarios.
Furthermore, \name achieves this high accuracy while maintaining impressive computational efficiency compared with AutoGluon, making it exceptionally suited for practical dynamic environments where both performance and speed are crucial. 
\textbf{(ii) Data Regression.}
Table~\ref{tab:reg} offers a comprehensive comparison of representative regression methods in the context of concept drift, focusing on movie rating prediction~\cite{IMDB-web}, a scenario typically characterized by evolving concepts.
\name excels in both mild and severe drift scenarios, maintaining consistent performance across MSE and $R^2$ metrics while demonstrating comparable efficiency.
While AutoGluon delivers the best results under mild drift conditions, its performance noticeably declines under severe drift and requires more than 40$\times$ computational time compared to \name.

\begin{table}[t]
   \small
    \centering
    \renewcommand{\arraystretch}{1}
    \caption{Performance of data classification on concept drift.}
    \label{tab:clf}
\resizebox{1\columnwidth}{!}{
        \begin{tabular}{l l c c c c c}
    \toprule[1.5pt]
  \multirow{2}{*}{Category} & \multirow{2}{*}{Method} & \multicolumn{2}{c}{Mild Drift}  & \multicolumn{2}{c}{Severe Drift} & Mean  \\
  & &  Acc$\uparrow$ &  F1$\uparrow$ & Acc$\uparrow$ & F1$\uparrow$ & Time(s)$\downarrow$ \\
    \midrule[0.5pt]
   \multirow{3}{*}{\makecell[l]{Classical  \\Non-linear Classifier}} & KNN  & 0.795 & 0.591 & 0.586 & 0.379 & \textbf{\underline{1.469}} \\
     & RandomForest  & 0.921 &  0.891 & 0.621 & 0.334 & 8.893\\
    & MLP  & 0.852 &0.585 & 0.676 & \textbf{0.496} & 15.798\\
     \midrule[0.1pt]
    \multirow{3}{*}{GBDT Classifier} & XGBoost &  0.905 & 0.896  & 0.596 & 0.385 & 55.681\\
     & LightGBM & 0.870 & 0.727 & 0.595 & 0.377 & 16.765\\
    & CatBoost & 0.918  &  0.906 & 0.607 & 0.368 & 14.077 \\
     \midrule[0.1pt]
   AutoML System & AutoGluon & \textbf{\underline{0.936}} & \textbf{0.908} & \textbf{0.679} & 0.441 & 85.183\\
    \midrule[0.1pt]
   Ours & \name & \textbf{0.932} & \textbf{\underline{0.920}} & \textbf{\underline{0.826}} & \textbf{\underline{0.632}} & \textbf{8.377} \\
    \bottomrule[1.5pt]
        \end{tabular}
    }
     \vspace{-2mm}
\end{table}

\begin{figure}[t]
    \centering 
\includegraphics[width=1\linewidth]{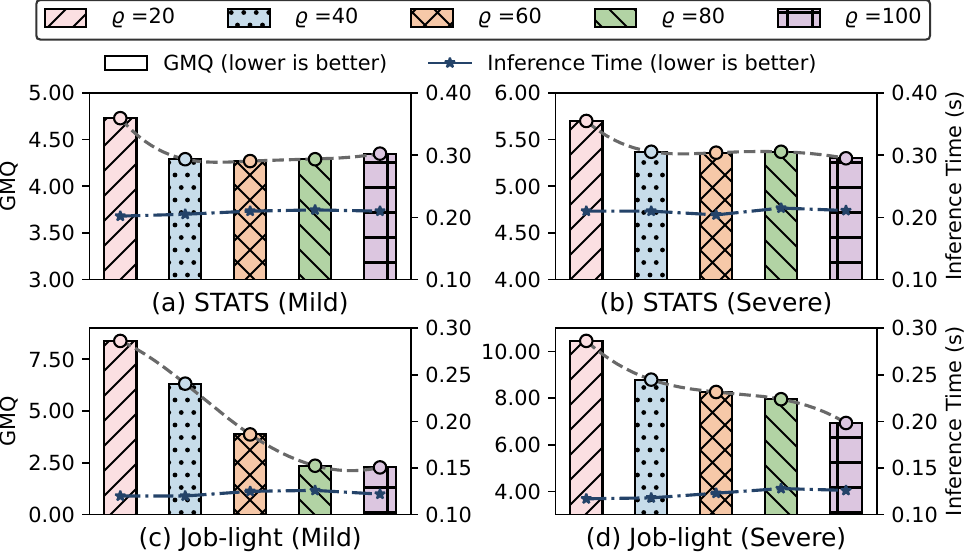} 
    \caption{Sensitivity analysis of the queue size $\varrho$.
    }
    \label{fig:queue_size}
\end{figure}

\subsection{Ablation Study}\label{sec:appendix-ablation}

\noindent \textbf{Effects of Queue Size in Context Memory.}
We further analyze the sensitivity of \name to the critical hyperparameter $\varrho$, the size of queues in context memory, across various benchmarks and dynamic scenarios, as depicted in~\cref{fig:queue_size}. 
The results confirm that increasing the queue size contributes to performance enhancements without escalating system latency, owing to embedding cache optimization.
Initially, performance improves significantly with an increase in queue size but eventually plateaus, indicating diminishing returns.
Notably, an oversized queue size may introduce information redundancy, potentially leading to a performance decline.
For instance, increasing the queue size to 100 results in a minor deterioration in the STATS benchmark's mild drift scenario.
In summary, the optimal queue size $\varrho$ should be tailored based on the complexity of the data to balance performance gains against the risk of redundancy, in order to optimize the model's efficacy in dynamic environments.

\begin{table}[t]
   \small
    \centering
    \renewcommand{\arraystretch}{1.2}
    \caption{Performance of data regression on concept drift.}
    \label{tab:reg}
    \resizebox{1\columnwidth}{!}{
        \begin{tabular}{l l c c c c c}
    \toprule[1.5pt]
  \multirow{2}{*}{Category} & \multirow{2}{*}{Method} & \multicolumn{2}{c}{Mild Drift}  & \multicolumn{2}{c}{Severe Drift} &Mean\\
  & &  MSE$\downarrow$ & $R^2$$\uparrow$ & MSE$\downarrow$ & $R^2$$\uparrow$ &Time(s)$\downarrow$ \\
    \midrule[0.5pt]
   \multirow{2}{*}{\makecell[l]{Classical \\ Method}} & SVR  & 0.591 & 0.230 & 0.691 & 0.210 & \textbf{0.081} \\
   & MLP & 8.762 & -10.418 & 28.355 & -49.003 & 10.259\\
   \midrule[0.1pt]
    \multirow{3}{*}{\makecell[l]{Tree-based \\Method}}  & DecisionTree  & 0.557 & 0.231 & 0.652 & 0.198 & \textbf{\underline{0.068}} \\
    & RandomForest & 0.315 & 0.570 & 0.458 & 0.475 & 0.942\\
    & GradientBoosting &  0.325 & 0.577 & \textbf{0.396} & 0.487 &0.355\\
    \midrule[0.1pt]
   AutoML System & AutoGluon & \textbf{\underline{0.267}} & \textbf{\underline{0.682}} & 0.399 & \textbf{0.632} & 27.438\\
   \midrule[0.1pt]
   Ours & \name & \textbf{0.271} & \textbf{0.647} & \textbf{\underline{0.388}} & \textbf{\underline{0.647}} & 0.681\\
    \bottomrule[1.5pt]
        \end{tabular}
    }
\end{table}

\begin{figure}[t]
    \centering 
\includegraphics[width=1\linewidth]{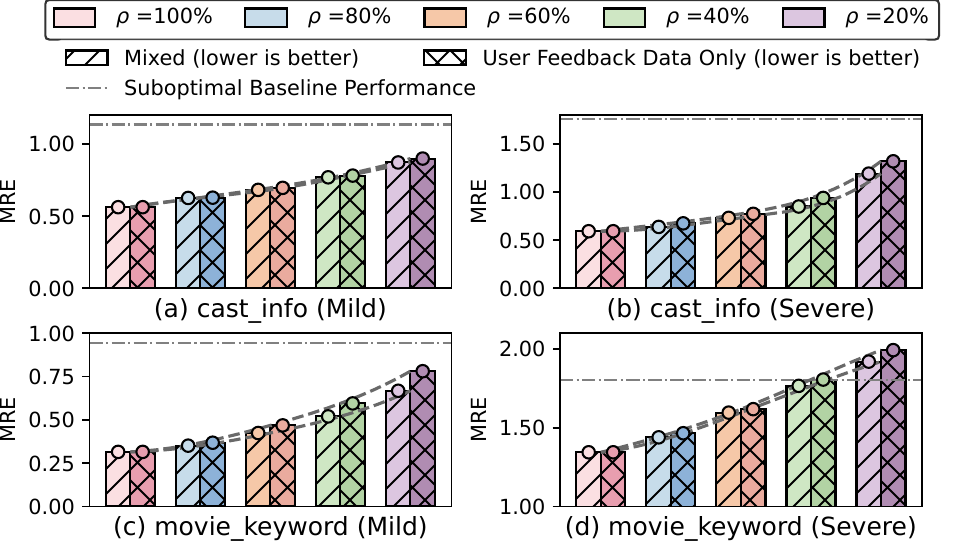} 
    \caption{Sensitivity analysis of the user feedback $\rho$.
    }
    \label{fig:user_feedback}
    \vspace{-2mm}
\end{figure}

\noindent \textbf{Effects of User Feedback.}
To delve into the adaptability of \name in user-oriented tasks, we 
evaluate how varying proportions of user feedback data $\rho$ within queues affect model performance.
We use drifted data with ground-truth outputs to simulate user-customized feedback data, assessing the model’s conformity to user-specific requirements.
Specifically, the queues comprise a certain proportion of user feedback data combined with the model's recent input-output pairs. 
We maintain the queue size at 80 and vary the proportion of user feedback data.
The results in~\cref{fig:user_feedback}, demonstrate that increasing the proportion $\rho$ within a fixed queue size significantly enhances model performance, confirming the model's ability to be customized by users.
To further explore the impact of integrating recent model interactions into the queue on performance,
we conduct comparative experiments using only user feedback data.
We observe that mixed queues outperform those containing solely user feedback. 
Additionally, integrating recent model data mitigates performance decline as the proportion $\rho$ of user feedback decreases. 
Still, we advise against setting $\rho$ too low due to the risk of introducing noise.
\blue{It is noteworthy that 
\name surpasses the suboptimal model DDUp at most times even with very low $\rho$,} underscoring \name's capability in user-oriented applications.

\begin{figure}[t]
    \centering
    \vspace{2mm}
\includegraphics[width=1\linewidth]{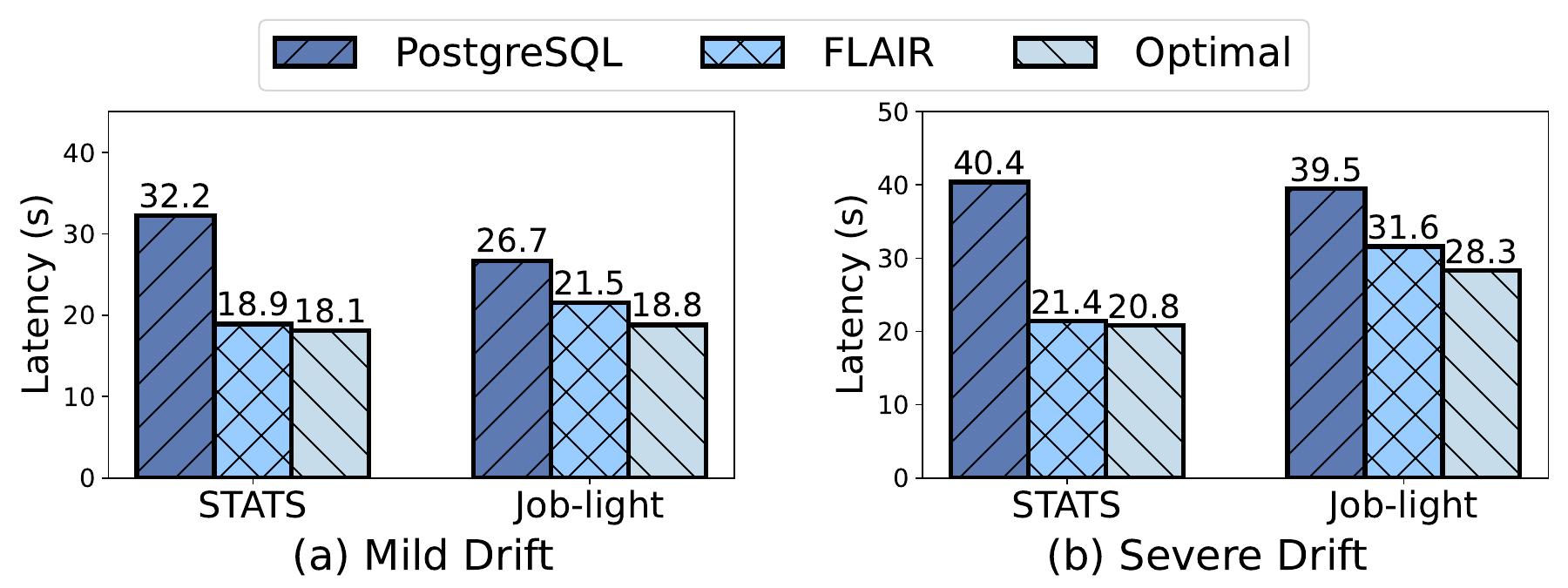} \vspace{-4mm}
    \caption{Comparison of query execution latency.
    }
    \label{fig:execution_time}
\end{figure}

\icmlblue{
\subsection{\name in Action}
}
\label{sec:exp_integration}
Given the observation from existing research~\cite{negi2021flow,marcus2021bao,li2023alece} that a smaller Q-error does not necessarily reduce execution times, we extend our investigation by integrating \name into PostgreSQL to assess its efficacy in a full-fledged database system. 
We evaluate the latency measured as execution time per query on the test set of STATS and JOB-light.
As in a recent work~\cite{li2023alece}, we 
substitute PostgreSQL's default cardinality estimator with \name. 
Specifically, PostgreSQL uses the cardinality estimated by \name to generate the execution plan for each query in the benchmarks.
The optimal baseline is established by replacing PostgreSQL's built-in estimations with ground-truth cardinalities.
As depicted in Figure~\ref{fig:execution_time}, \name achieves latency that approaches the optimal level based on ground-truth cardinality. 
Compared to PostgreSQL’s built-in cardinality estimator, \name accelerates query execution by up to 1.9$\times$. This superiority is even more significant in severe drift scenarios.
\section{Related Work}
\label{sec:appendix-related work}
\subsection{Advances and Challenges of AI×DB}
\label{sec:appendix-rw-aidb}
Database systems are increasingly embracing artificial intelligence (AI), spurring the development of AI-powered databases (AI×DB)~\cite{ooi2024neurdb,
zhu2024pilotscope,li2021ai,mcgregor2021preventing}.
This fusion marks a new era for database systems, in which
AI functionalities are incorporated to enhance the overall system performance and usability.
Consequently,
advanced models such as deep neural networks (DNNs) and large language models (LLMs) are increasingly being integrated into 
database systems
and applications, which has 
improved
database management
such as
database tuning~\cite{lao2024gptuner,huang2024llmtune,trummer2022db}, cardinality and selectivity estimation~\cite{lee2024asm,kurmanji2023detect,li2023alece,hilprecht2019deepdb}, and indexing~\cite{zhang2024making,li2020lisa,li2023dili,gao2023lmsfc,sun2023learned,zhang2024hyper}.
Recent work~\cite{zeighami2024theoretical} 
presents a theoretical foundation for developing machine learning approaches in database systems.
However,
unlike the data that AI models have been designed for, online transactional processing (OLTP) data is dynamic in nature and
such dynamicity affects the robustness of models. 
Indeed, the phenomenon of \textit{concept drift}, where the underlying data distributions and relations shift, 
remains a critical challenge. 
\green{
In this study, our goal is to provide a solution for addressing \textit{concept drift} in databases, ensuring both accuracy and sustainability in dynamic environments.
}
%

\subsection{Model Adaptation in Concept Drift}
\label{sec:appendix-rw-concept-drift}
Variations in data critically affect the efficacy of AI-powered database systems, also known as learned database systems.
Such discrepancies between training data and those encountered post-deployment significantly degrade system performance, challenging model reliability in dynamical environments for the practical deployment~\cite{negi2023robust,zeighami2024theoretical}.
Recent cutting-edge machine learning paradigms such as transfer learning~\cite{jain2023data,kurmanji2023detect, kurmanji2024machine,ying2018transfer}, active learning~\cite{ma2020active,li2022warper,lampinen2024passive}, and multi-task learning~\cite{kollias2024distribution,wu2021unified,hu2024revisiting} have been employed to mitigate challenges of concept drift in AI-powered database systems.
Notably, Kurmanji et al. utilize knowledge distillation, guided by loss-based out-of-distribution data detection for handling data insertions~\cite{kurmanji2023detect}, and explore transfer learning for machine unlearning to address data deletions in database systems~\cite{kurmanji2024machine}.
Additionally, reinforcement learning (RL) has been used to strategically reduce the high costs of data collection by allowing an RL agent to selectively determine which subsequent queries to execute with a more targeted fashion~\cite{zhang2019end,hilprecht2020learning,zheng2024robustness,wang2023adopt}.
These strategies, while aimed at 
improving generalization in fluctuating environments, inherently face critical issues due to their 
requirements for data recollection and model retraining.
For instance, optimizing query performance necessitates executing 
numerous query plans, a process that is computationally intensive and significantly extends execution time~\cite{wu2021unified,hilprecht2021one,li2022warper}.
The need for repetitive executions, whenever new concepts are detected, further compounds the operational challenges.

Inspired by large language models (LLMs), 
zero-shot learning has been employed to enhance model adaptability to dynamic environments and generalize across different tasks~\cite{hilprecht2021one,zhou2023d,urban2023omniscientdb}. 
While this approach is theoretically promising, it faces practical challenges, as pre-training or fine-tuning large foundation models still requires substantial real-world data collection. 
Additionally, the quality and relevance of training data to actual workloads remain uncertain until deployment, making post-deployment performance unpredictable. 
Further, existing methods struggle to keep pace with real-time evolving concepts and overlook inter-query relations, which compromises their effectiveness.
To fundamentally address these challenges, we propose a fresh perspective on online adaptation for database systems that supports on-the-fly in-context adaptation to evolving concepts without unnecessary data collection or retraining, ensuring unparalleled effectiveness and efficiency in operational settings.

\section{Conclusions}
\label{sec:conclusion}
With frequent updates, 
the data in the database evolves,
resulting in concept drift.
Learned database operations
are susceptible to concept drift, and may suffer significant prediction accuracy losses.
This paper presents a 
novel online adaptation framework called \name, which can adapt the in-database predictive model to evolving concepts automatically without cumbersome data recollection and model retraining. 
\name performs Bayesian meta-training using abundant synthetic data sampled from dynamic task distributions. After meta-training, it generates adapted predictions by prompting the model with contextual information regarding the current concept.
%
Extensive experiments across various database operations demonstrate that \name is effective, efficient, and transferable in dynamic data systems.
\ignore{
Extensive experiments demonstrate that \name accelerates adaptation speed by $5.2\times$ and decreases the geometric mean of Q-error by 22.5\% compared with the advanced baseline method in cardinality estimation. 
By integrating \name with PostgreSQL, the query execution latency has been reduced by up to $1.9\times$.
}



\section*{Impact Statement}\label{sec:impact_statement}
This paper showcases advances in the integration of Machine Learning within data systems, presenting a framework that enhances the reliability and usability of Machine Learning in real-world dynamic environments.
By facilitating various database operations, the proposed framework encourages broader adoption in practice, potentially reducing operational costs and energy consumption.
There can be various societal consequences, including enhancing decision-making in data-reliant sectors such as healthcare, finance, and public services, and enabling businesses to achieve significant cost reductions through optimized operations and minimized manual intervention.

\ignore{
Authors are \textbf{required} to include a statement of the potential broader impact of their work, including its ethical aspects and future societal consequences. 
This statement should be in an unnumbered section at the end of the paper (co-located with Acknowledgements -- the two may appear in either order, but both must be before References), and does not count toward the paper page limit. 
In many cases, where the ethical impacts and expected societal implications are those that are well established when advancing the field of Machine Learning, substantial discussion is not required, and a simple statement such 
as the following will suffice:

``This paper presents work whose goal is to advance the field of Machine Learning. There are many potential societal consequences of our work, none of which we feel must be specifically highlighted here.''

The above statement can be used verbatim in such cases, but we encourage authors to think about whether there is content which does warrant further discussion, as this statement will be apparent if the paper is later flagged for ethics review.
}


\nocite{langley00}


\bibliographystyle{ACM-Reference-Format}
\bibliography{reference}



\end{document}